\documentclass[fleqn,usenatbib]{mnras}

\usepackage{newtxtext,newtxmath}
\usepackage[T1]{fontenc}
\usepackage{ae,aecompl}

\usepackage{xcolor}

\usepackage{graphicx}	% Including figure files
\usepackage{amsmath}	% Advanced maths commands
\usepackage{amssymb}	% Extra maths symbols

\title[Planet formation in an MRI accretion disc]{Close-in Super-Earths:\\ The first and the last stages of planet formation in an MRI-accreting disc}

\author[M. R. Jankovic et al.]{
Marija R. Jankovic,\thanks{E-mail: m.jankovic16@imperial.ac.uk}
James E. Owen and
Subhanjoy Mohanty
\\
Astrophysics Group, Imperial College London, Blackett Laboratory, Prince Consort Road, London SW7 2AZ, UK
}

\date{Accepted XXX. Received YYY; in original form ZZZ}

\pubyear{2018}

\begin{document}
\label{firstpage}
\pagerange{\pageref{firstpage}--\pageref{lastpage}}
\maketitle

% Abstract of the paper
\begin{abstract}
We explore \textit{in situ} formation and subsequent evolution of close-in super-Earths and mini-Neptunes. We adopt a steady-state inner protoplanetary gas disc structure that arises from viscous accretion due to the magneto-rotational instability (MRI). We consider the evolution of dust in the inner disc, including growth, radial drift and fragmentation, and find that dust particles that radially drift into the inner disc fragment severely due to the MRI-induced turbulence. This result has two consequences: (1) radial drift of grains within the inner disc is quenched, leading to an enhancement of dust in the inner regions which scales as dust-to-gas-mass-flux-ratio at $\sim$\,1\,AU; (2) however, despite this enhancement, planetesimal formation is impeded by the small grain size. Nevertheless, assuming that planetary cores are present in the inner disc, we then investigate the accretion of atmospheres onto cores and their subsequent photoevaporation. We then compare our results to the observed exoplanet mass-radius relationship. We find that: (1) the low gas surface densities and high temperatures in the inner disc reduce gas accretion onto cores compared to the minimum mass solar nebula, preventing the cores from growing into hot Jupiters, in agreement with the data; (2) however, our predicted envelope masses are still typically larger than observed ones. Finally, we sketch a qualitative picture of how grains may grow and planetesimals form in the inner disc if grain effects on the ionization levels and the MRI and the back-reaction of the dust on the gas (both neglected in our calculations) are accounted for. 
\end{abstract}

% Select between one and six entries from the list of approved keywords.
% Don't make up new ones.
\begin{keywords}
protoplanetary discs  --- 
planets and satellites: formation
\end{keywords}

%%%%%%%%%%%%%%%%%%%%%%%%%%%%%%%%%%%%%%%%%%%%%%%%%%

%%%%%%%%%%%%%%%%% BODY OF PAPER %%%%%%%%%%%%%%%%%%

\section{Introduction} \label{sec:intro}

Recent advances in exoplanet detection, led primarily by the {\it Kepler} mission, have uncovered several new classes of exoplanets \citep[e.g.][]{Borucki2011,Batalha2013}. These are
the close-in super-Earths and mini-Neptunes, planets with radii of 1\,--\,4\,R$_{\oplus}$ and periods of up to 100 days, and they are found to be abundant around solar and sub-solar mass stars \citep[e.g.][]{Fressin2013, Dressing2015}. How (and where in their parent protoplanetary discs) these planets form is a subject undergoing intense study.

One suggestion is that these super-Earths/mini-Neptunes form at larger separations, as more solids are potentially available outside the ice line, and then migrate inwards through the disc \citep[e.g.][]{IdaLin,Kley2012, Cossou2013, Cossou2014}. However, the migration scenario predicts that planets in multi-planet systems should typically end up in mean motion resonances, whereas such orbital resonances are rare among the \textit{Kepler} planets \citep[e.g.][]{Baruteau2014,Winn2015}. Although several mechanisms to either break, or prevent the capture into, resonances have been explored \citep[e.g.][]{GoldreichSchlichting2014,Izidoro2017,Liu2017a}, this discrepancy has not yet been fully resolved. Moreover, the radius distribution of the \textit{Kepler} planets, shaped by atmospheric photoevaporation, appears consistent with the planetary cores having a rock/iron (Earth-like) composition \citep{Owen2017}, implying formation inside the ice line and arguing against significant migration.

An alternative scenario is that these planets form \textit{in situ}, close to their present orbits. In this case, planetary cores form in the inner protoplanetary disc. They can still be subject to planet-disc interactions and the two are not mutually exclusive. If the cores are subject to the fast type I migration, the innermost planet could stall at the inner disc edge \citep{Ogihara2015, Masset2006}, explaining why the observed period distribution of the innermost planet peaks around $\sim$\,10 days (\citealt{Mulders2018}; see also \citealt{Mulders2015}, \citealt{Lee2017b}). Alternatively, type I migration could be suppressed if the surface density profile is flat or has a positive slope in the inner disc \citep{Ogihara2018}, or stalled by the core opening a gap \citep[e.g.][]{Hu2016}. In fact, there is evidence suggesting that the \textit{Kepler} planets could have been massive enough to open gaps in the inner disc \citep{Wu2018}.

If the close-in planets do form \textit{in situ}, a large amount of solids is necessary in the inner protoplanetary disc compared to the amount in the minimum mass solar nebula \citep{Chiang2013}. These solids may be delivered to the inner disc from the outer disc prior to planet formation \citep{Hansen2012, Hansen2013, Chatterjee2014}, through the radial drift of pebbles and rocks \citep{Weidenschilling1977, Takeuchi2002, ArmitageLectureNotes}. %Specifically, the gas pressure in circumstellar discs is expected to decrease outwards, and the resulting outward thermal pressure gradient makes the gas revolve at sub-Keplerian velocities. As dust grains grow more massive through coagulation, however, their large inertia makes them immune to the thermal pressure, so pebbles and rocks tend to orbit at Keplerian velocities. Hence these solids experience a headwind from the slower moving gas (gas drag) which robs them of angular momentum, causing them to radially drift inwards towards the star and the inner disc} \citep{Weidenschilling1977,Takeuchi2002,ArmitageLectureNotes}. 
The growth of dust grains in the outer disc and their subsequent radial drift inwards have been confirmed by observations \citep[e.g.][]{Panic2009, Andrews2012, Isella2012, Rosenfeld2013, Powell2017}.

To create a dust-rich inner disc in which to form planets, the radial drift of dust particles needs to be stopped or slowed down. The radial drift of particles in the Epstein drag regime slows down closer to the star in conventional disc models, and this can concentrate dust in the inner disc to some extent \citep{YoudinShu2002, YoudinChiang2004, Birnstiel2010, Birnstiel2012, Drazkowska2016}.
%for small particles of constant size \citep{YoudinShu2002, YoudinChiang2004}, and similarly for particles whose size is set by collisional fragmentation due to turbulence \citep{Birnstiel2010, Birnstiel2012, Drazkowska2016}
Another way to halt the radial drift and enrich the inner disc with dust is to trap the solids inside an axisymmetric local gas pressure maximum that is expected to form if the accretion in the inner disc is driven by the magneto-rotational instability \citep[MRI;][]{Kretke2009, Dzyurkevich2010, Drazkowska2013, Chatterjee2014}. A gas pressure maximum acts as a trap for the marginally-coupled solids as the gas inwards of the pressure maximum is super-Keplerian, reversing the direction of the radial drift \citep[e.g.][]{Pinilla2012}. Furthermore, in a steady-state disc accreting due to the MRI a pressure maximum forms at the boundary between the thermally ionized innermost disc in which the MRI-induced viscosity is high, and the low-viscosity \textit{dead zone} in which the MRI is suppressed due to low ionization levels \citep{Gammie1996}. This local pressure maximum is expected to form at few tenths of AU around solar and sub-solar stars \citep{Chatterjee2014}, which is consistent with the orbital distances of the close-in super-Earths and mini-Neptunes.

\citet{Mohanty2017arxiv} presented a semi-analytic steady-state inner disc model in which the disc structure, thermal ionization and the viscosity due to the MRI were determined self-consistently. The location of the pressure maximum inferred from this model is similarly in general agreement with the orbital distances of the close-in planets. An important insight from the \citet{Mohanty2017arxiv} models of the inner disc is that in a steady state they predict gas surface densities that are considerably lower than those of the minimum mass solar nebula. This is not surprising as the minimum mass solar nebula simply extrapolated the surface density to small separations, whereas in reality the shrinking size of the dead-zone results in more efficient angular momentum transport, hence lowering the surface densities towards smaller separations. 

The atmospheres of many of the close-in super-Earths and mini-Neptunes must be H/He dominated \citep[e.g.][]{JontofHutter2016} and they typically make up 0.1\,--\,10\,\% of their total mass \citep{Lopez2014, Wolfgang2015}. Thus, they are considerably larger than the atmospheres of the planets in the inner Solar system. Outgassing of hydrogen from a rocky core is not sufficient to explain the majority of these atmospheres \citep{Rogers2011}. Thus these atmospheres are most likely composed of gas accreted from the protoplanetary disc after the formation of a solid core. If so, these atmospheres are formed steadily through core accretion.

\citet{Lee2014} \citep[see also][]{Lee2015,Lee2016,Lee2017} argue that core accretion is so efficient that the key concern is how to stop the super-Earth cores from undergoing runaway accretion and becoming gas giants \citep{Mizuno1980}. This led \citet{Lee2016} to suggest that super-Earth/mini-Neptune formation occurred in gas-poor ``transition discs'', during the final short-lived phase of disc dispersal. The requirement for a gas-poor inner disc raises the question if planet formation in the gas-poor inner disc arising due to steady-state MRI accretion could be a desirable scenario and a possible alternative to the \citet{Lee2016} proposal of atmospheric accretion during disc dispersal.

%In this scenario the gaseous envelope is connected to the disc and the principal model hypothesises it is in a quasi-hydrostatic equilibrium and contracts as it cools, allowing more gas to be accreted \citep[e.g.][]{Rafikov2006,Lee2014,Lee2015}.

%The abundance of close-in, gas-rich planets raises the question of whether the planetary cores can accrete sufficiently large amounts of gas if they formed in an inner disc that is not only gas-poor, but also hot enough to ionize alkali elements in order to support the MRI. Various studies have shown, in application to an inner disc depleted in gas due to disc dispersal, that planets can still accumulate large amounts of gas \citep{Ikoma2012,Inamdar2015,Lee2016}.

In this paper, we examine the possibility of the \textit{in situ} formation of the close-in planets in the inner disc structure arising from MRI-driven accretion, obtained using the self-consistent model of \citet{Mohanty2017arxiv}. First, in section \ref{sec:dust}, we examine the evolution of dust in the inner disc and discuss the possibility of planetesimal formation. In section \ref{sec:atmospheres}, we calculate the atmospheres that super-Earth and mini-Neptune cores can accrete in the gas-poor inner disc implied by the MRI, and then evolve them forward in time, in order to compare our calculations to the data.

An important caveat to our dust calculations is that we evolve the dust assuming the gas profile is fixed in time. Dust grains can act to suppress the MRI by lowering the coupling between the gas and the magnetic field \citep[e.g.][]{Sano2000,Ilgner2006}, and could thus significantly alter the gas disc structure. Moreover, at high dust-to-gas ratios, which we show can be achieved in the inner disc, dust becomes dynamically important and the dynamical back-reaction on the gas should be taken into account. In section \ref{sec:discussion}, we qualitatively discuss these effects and how they might influence our results. We shall address the self-consistent feedback of dust enhancement on the gas disc structure in subsequent studies.

\section{Dust evolution} \label{sec:dust}
\begin{figure}
\centering
\includegraphics[height=0.9\textheight]{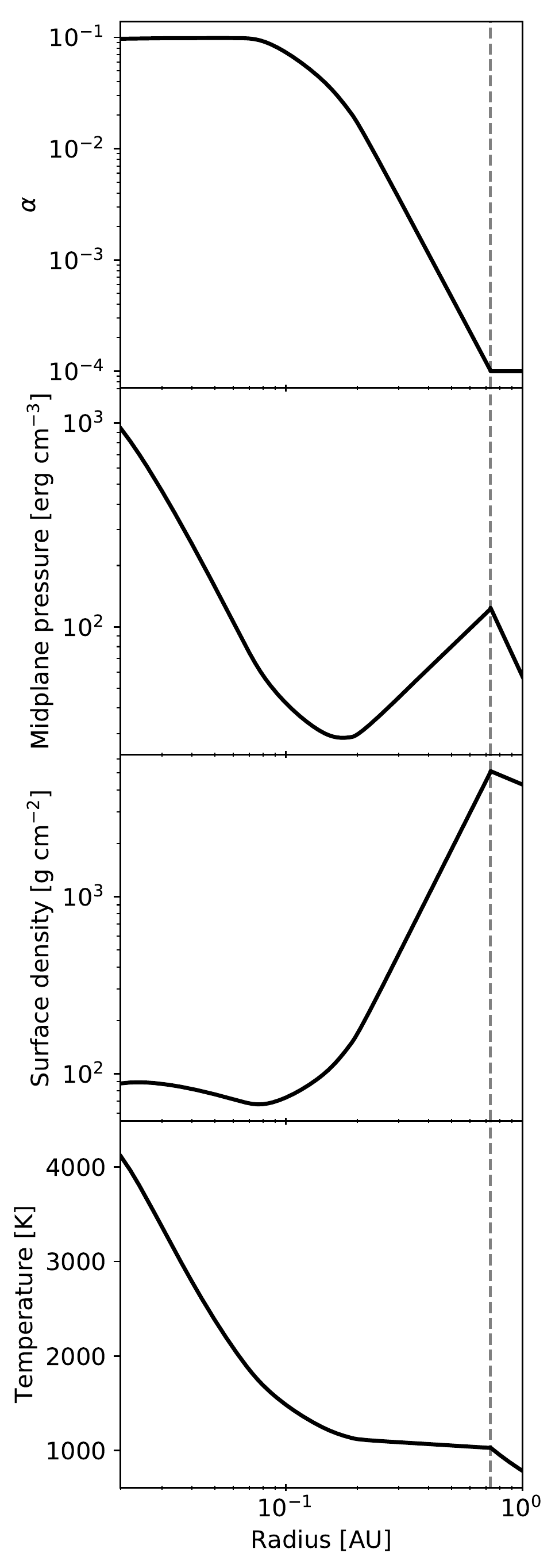}
\caption{\label{fig:gas_structure}
Gas disc structure from the steady-state model of \citet{Mohanty2017arxiv} for $M_{\ast}=1$\,M$_\odot$, $R_{\ast}=2.34$\,R$_\odot$, $\dot{M}_{\rm g} = 10^{-8}$\,M$_\odot$\,$\rm{yr}^{-1}$ and $\alpha_{\rm DZ}=10^{-4}$. From top to bottom: $\alpha$ parameter, midplane pressure, surface density and temperature, as functions of radius. Location of the local gas pressure maximum due to the MRI is indicated by the vertical dashed line.}
\end{figure}

We consider the evolution of the dust, including growth, fragmentation and radial drift, in a steady-state gas disc which is viscously accreting due to the MRI. The structure of the gas disc (gas surface density, temperature, pressure and viscosity) is obtained from the steady-state models of the inner protoplanetary gas disc calculated by \citet{Mohanty2017arxiv}. In these models the viscosity ($\alpha$ parameter) is determined self-consistently with the disc structure \citep{ShakuraSunyaev1973}, thermal ionization and MRI criteria \citep{BaiStone2011,Bai2011}. The parameters of the model are the stellar mass and radius ($M_{\ast}$, $R_{\ast}$); steady state accretion rate ($\dot{M}_{\rm g}$); and minimum viscosity of the gas due to purely hydrodynamical effects ($\alpha_{\rm DZ}$). The minimum viscosity $\alpha_{\rm DZ}$ is the assumed value of $\alpha$ inside the MRI-dead zones; it is a minimum in the sense that such hydrodynamical effects are assumed to dominate over the MRI-induced turbulence if the MRI implies a viscosity lower than $\alpha_{\rm DZ}$. In this work we primarily use $M_{\ast}=1$\,M$_\odot$, $R_{\ast}=2.33$\,R$_\odot$, $\dot{M}_{\rm g} = 10^{-8}$\,M$_\odot$\,$\rm{yr}^{-1}$ and $\alpha_{\rm DZ}=10^{-4}$. The disc structure for these parameters is shown in Fig. \ref{fig:gas_structure}. The local gas pressure maximum is at an orbital distance of $\sim$\,0.7\,AU, and temperature and surface density at that location are $\sim$\,1000\,K and $\sim$\,5000\,g\,cm$^{-2}$, respectively. Outwards from the pressure (and the surface density) maximum the MRI is suppressed and $\alpha=\alpha_{\rm DZ}$.

Throughout this paper we use the above fiducial values for the disc parameters. Here we briefly describe how the disc structure depends on these parameters. For a higher gas accretion rate $\dot{M}_{\rm g}$ the radial $\alpha$ profile is roughly an outward-translated version of the one shown in the top panel of Fig. \ref{fig:gas_structure}, and inward-translated for a smaller $\dot{M}_{\rm g}$. The radial location of the local gas pressure maximum scales with the accretion rate approximately as $\dot{M}_{\rm g}^{1/2}$. For a higher $\alpha_{\rm DZ}$ the $\alpha$ falls to this value closer to the star, and vice versa, but the value of $\alpha$ as a function of orbital distance remains almost the same otherwise; as a result, the radial location of the local pressure maximum scales with the minimum dead-zone viscosity as $\alpha_{\rm DZ}^{-1/4}$. Furthermore, inwards of the pressure maximum the temperature has to be sufficiently high for thermal ionization of potassium to support the MRI and so it is always larger than 1000\,K, regardless of the exact choice of $\dot{M}_{\rm g}$ and $\alpha_{\rm DZ}$. For a steady-state vertically-isothermal $\alpha$-disc the accretion rate is $\dot{M}_{\rm g}=3 \pi c_{\rm s}^2 \alpha \Sigma_{\rm g} / \Omega$ (ignoring an additional factor which depends on the boundary condition at the inner disc edge, and which becomes unimportant far away from the edge). Further assuming (as is approximately the case) that the temperature at the location of the pressure maximum is constant regardless of the disc parameters, it follows that the maximum gas surface density approximately depends on the disc parameters\footnote{Derivation of the scalings from the Shakura-Sunyaev equations takes into account the small correction due to the dependence of the temperature on the disc parameters and yields that the maximum surface density depends on the disc parameters as $\dot{M}_{\rm g}^{3/10}$ and $\alpha_{\rm DZ}^{-13/20}$. These small corrections are omitted here for simplicity.}as $\dot{M}_{\rm g}^{1/4}$ and $\alpha_{\rm DZ}^{-5/8}$.

As we neglect the dynamical back-reaction of the dust on the gas and the effect of the dust on the MRI, the local evolutionary timescales are considerably shorter than the Myr timescale on which the accretion rate will evolve. Thus, the structure of the gas disc is held fixed, i.e. not evolved in time. Note also that our gas disc model considers MRI-driven accretion only, and magnetic winds, if present, can also affect the inner gas disc structure \citep[e.g. reduce the gas surface density compared to the minimum mass solar nebula even in the absence of the MRI,][]{Suzuki2016}.

\subsection{Methods} \label{sec:dust_methods}

The dust particle size distribution is evolved using the two-population model of \citet{Birnstiel2012}. The dust surface density $\Sigma_{\rm d}$ is evolved using the advection-diffusion equation
\begin{equation}
    \frac{\partial \Sigma_{\rm d}}{\partial t} + \frac{1}{r} \frac{\partial}{\partial r} \left[ r \left( \Sigma_{\rm d} \bar{u} - D_{\rm gas} \Sigma_{\rm g} \frac{\partial}{\partial r}\left( \frac{\Sigma_{\rm d}}{\Sigma_{\rm g}} \right) \right) \right] = 0 \,,
	\label{eq:advection-diffusion}
\end{equation}
where $r$ is the cylindrical radius, $\bar{u}$ is the dust advection velocity, $D_{\rm gas}$ is the gas diffusivity and $\Sigma_{\rm g}$ is the gas surface density. 

The dust advection velocity is a sum of the velocities due to advection with the accreting gas and radial drift. For particles with Stokes number St$_i=\pi \rho_{\rm s} a_i/2\Sigma_{\rm g}$ (with $\rho_{\rm s}$ the internal density of the dust and $a_i$ the particle size), and adopting the terminal velocity approximation \citep[e.g.][]{Takeuchi2002}, the dust velocity is given by
\begin{equation}
	u_i = \frac{1}{1+\textrm{St}_i^2} u_{\rm gas} + \frac{2}{\textrm{St}_i+\textrm{St}_i^{-1}} u_{\rm drift} \,,
	\label{eq:advection_velocity}
\end{equation}
where 
\begin{equation}
	u_{\rm drift} = \frac{c_{\rm s}^2}{2 v_{\rm K}} \frac{\textrm{d\,ln} P}{\textrm{d\,ln} r} \,,
\end{equation} 
with $c_{\rm s}$ the speed of sound, $v_{\rm K}$ the Keplerian velocity and $P$ the midplane gas pressure. Small particles (St\,$\ll$\,1) move with the gas, and larger particles can move faster or slower than the gas, depending on the sign of the pressure gradient.

In the \citet{Birnstiel2012} model the dust surface density $\Sigma_{\rm d}$ and dust advection velocity $\bar{u}$ are the sum and the mass-weighted average, respectively, of the surface density and velocity of two populations of particles: small monomer-sized particles ($a_{\rm 0}=1$\,$\mu$m) and large particles ($a_{\rm 1}$). The size of the large particles evolves in time and space. At first, small dust grains grow. Then, at each radius the size of the large particles is set by whichever process yields the smallest grain size: radial drift ($a_{\rm drift}$), where grains larger than $a_{\rm drift}$ radially drift more quickly than they can grow; drift-fragmentation ($a_{\rm df}$), where grains larger than $a_{\rm df}$ fragment due to relative radial drift velocities; or turbulent fragmentation ($a_{\rm frag}$), where grains larger than $a_{\rm frag}$ fragment due to relative velocities induced by turbulence.

As we are interested in the innermost protoplanetary disc, inside the ice line, we set the bulk density of particles to $\rho_{\rm s}=3$\,g\,cm$^{-3}$, and the critical fragmentation velocity to $u_{\rm f}=1$\,m\,s$^{-1}$, based on experiments on collisions of silicate grains \citep{BlumMunch1993, Beitz2011, Schrapler2012, Bukhari_Syed2017} of similar size (the regime applicable in the \citealt{Birnstiel2012} model used here).\footnote{We note that simulations of grain collisions \citep{Meru2013} indicate that the critical fragmentation velocity could be significantly higher for porous grains than for compact ones, for a range of porosities that is not robustly covered by the above experiments. Now, in the fragmentation-limited regime particle size depends quadratically on $u_{\rm f}$. Thus, {\it if} porosity is important, and hence $u_{\rm f}$ is set to e.g. $10$\,m\,s$^{-1}$ instead, particle sizes (Stokes number) would be larger by a factor of 100, strongly affecting how coupled a particle is to the gas flow and how susceptible to radial drift. However, since particles also become less porous (to the point of becoming compact, and fragile) in a wide variety of conditions -- e.g., in collisions that result in coagulation (simulations by \citealt{Meru2013}, experiments by \citealt{Kothe2010}), collisions that result in bouncing \citep{Weidling2009} and collisions of larger grains with monomers \citep{SchraplerBlum2011} -- we opt to use the compact grain value of $u_{\rm f}=1$\,m\,s$^{-1}$.}

The viscosity parameter $\alpha$ due to the MRI-turbulence from our gas disc model is determined as a vertical average at each radius. We assume that this vertically-averaged $\alpha$ signifies the strength of turbulence that particles feel, which in turn determines the particle size due to turbulent fragmentation ($a_{\rm frag}$) and the radial turbulent mixing (diffusivity $D_{\rm gas}$). However, the viscosity (and the level of turbulence) can be different at the disc midplane compared to the upper layers of the disc, depending on where the non-ideal magnetohydrodynamic effects suppress the MRI. Dust tends to settle towards midplane, its scale-height being determined by the balance between gravitational settling and turbulent stirring \citep[e.g.][]{Youdin2007}. The use of a vertically-averaged $\alpha$ could thus be invalid in weakly turbulent regions. Nevertheless, we proceed with this assumption for ease of computation; we do check the robustness of our results by swapping the vertically-averaged $\alpha$ parameter for the midplane value in one run, and recover qualitatively the same results.

\subsection{Numerical procedure}
The advection-diffusion equation (\ref{eq:advection-diffusion}) is integrated using an explicit first order in time and second order in space finite element method. The advection term is integrated with an upwind scheme that adopts a van Leer flux limiter; the numerical scheme is described in detail by \citet{Owen2014}. The only modification we have made is the inclusion of the \citet{Birnstiel2012} dust evolution algorithm. Our simulations use 262 cells in the radial direction with a 0.002\,AU spacing inwards of 0.4\,AU and a 0.01\,AU spacing outwards. The inner boundary is set to 0.016\,AU, and the outer boundary is set to 1\,AU (i.e. outside the pressure maximum, but inside the ice line). The time-step is set with respect to the spatial resolution, the advection speed and the diffusion coefficient, so that it obeys the Courant-Friedrichs-Lewy condition. Following \citet{Birnstiel2012}, at each time step the size of the large particles is updated to the smallest of the four size limits ($a_{\rm growth}$, $a_{\rm drift}$, $a_{\rm df}$ or $a_{\rm frag}$). At the outer boundary we impose a constant dust accretion rate, which we vary in different runs. This is to mimic the fact that the dust flux is not a fixed quantity and can vary with time due to radial drift of dust from the outer disc \citep[e.g.][]{Birnstiel2012}. Dust particle size at the outer boundary is calculated self-consistently.

Furthermore, we neglect the effects of dust sublimation, and note that the temperature in our gas disc model exceeds the dust sublimation value ($\sim$$1500$\,K) only inwards of $0.1$\,AU.

\subsection{Results} \label{sec:dust_results}

\begin{figure}
\centering
\includegraphics[width=\columnwidth]{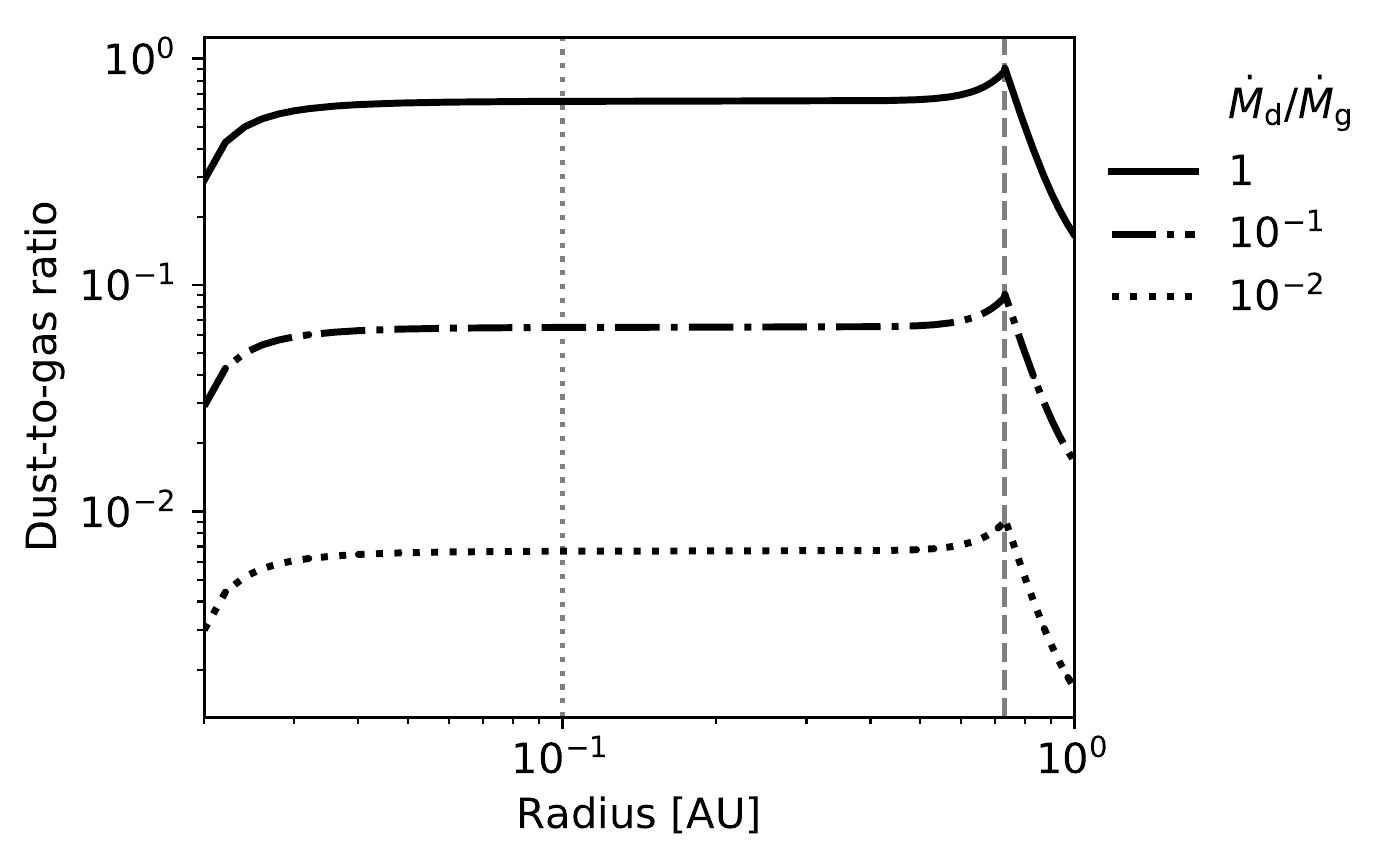}
\caption{\label{fig:dust_to_gas_ratio}
Dust-to-gas ratio $\Sigma_{\rm d}/\Sigma_{\rm g}$ as a function of radius after $0.2$\,Myr, for various dust accretion rates at the outer boundary, as indicated in plot legend. Location of the local gas pressure maximum due to the MRI is indicated by the vertical dashed line, and location of the dust sublimation line by the vertical dotted line.}
\end{figure}

Our simulations proceed as follows: initially, the dust-to-gas ratio is $0.01$ at all radii and all dust grains are monomers ($a=1$\,$\mu$m). We evolve the dust for $0.2$\,Myr, by which time it has reached steady state. The steady-state dust-to-gas ratio is shown in Fig. \ref{fig:dust_to_gas_ratio} as a function of radius for three different values of dust accretion rates $\dot{M}_{\rm d}$ at the outer boundary condition. For any given dust accretion rate, the steady-state dust-to-gas ratio is roughly constant inwards of the pressure maximum (indicated by the vertical dashed line), and it decreases outwards from the pressure maximum. There is only a moderate accumulation of dust at the pressure maximum, compared to the rest of the inner disc. This implies that the pressure maximum does not efficiently trap dust particles.

\begin{figure}
\centering
\includegraphics[width=\columnwidth]{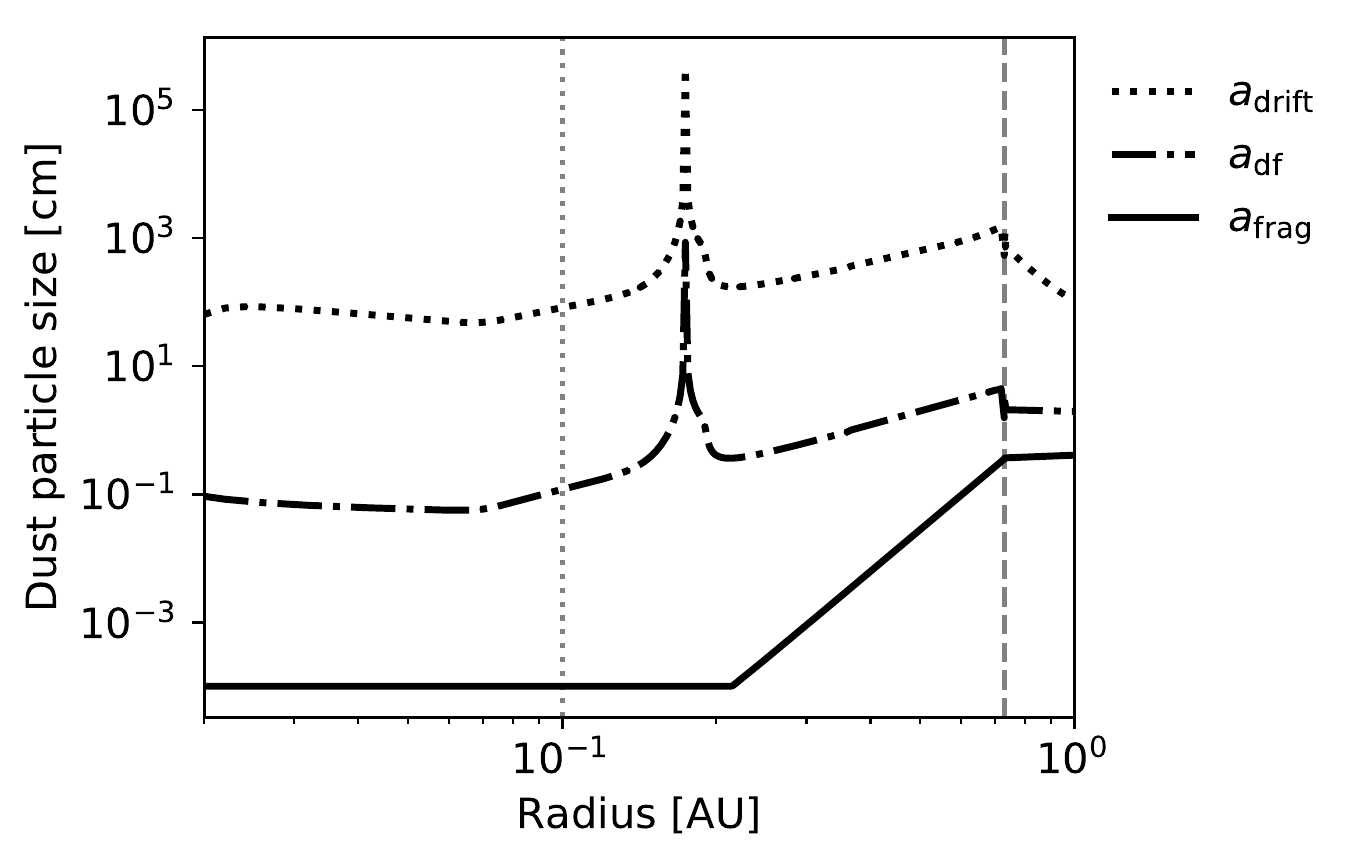}
\caption{\label{fig:dust_size}
Dust particle size limits due to radial drift ($a_{\rm drift}$), drift-fragmentation ($a_{\rm df}$), and turbulent fragmentation ($a_{\rm frag}$) as functions of radius after $0.2$\,Myr. Location of the local gas pressure maximum due to the MRI is indicated by the vertical dashed line, and location of the dust sublimation line by the vertical dotted line. Spikes in $a_{\rm drift}$ and $a_{\rm df}$ correspond to gas pressure extrema.}
\end{figure}

Essentially, the particles do not feel significant gas drag, and thus do not significantly feel the effect of the change in the sign of the pressure gradient inwards of the pressure maximum. This happens because, the MRI-induced turbulence causes fragmentation, resulting in small particles. Fig. \ref{fig:dust_size} shows the three dust size limits (due to radial drift, drift-fragmentation and turbulent fragmentation) as functions of radius, calculated in steady state. The smallest of the three ($a_{\rm frag}$), due to turbulent fragmentation, sets the size of the population of large particles in these simulations, which dictates the evolution of dust overall. Particle size is thus limited to only a few millimeters near the pressure maximum (indicated by the vertical dashed line), and the particles are monomer-sized in the innermost disc.

The particle size determines, through the Stokes number, how coupled the dust is to the gas. Thus, the particle size determines to what extent the particles move with the accreting gas towards the star and also by how much they are slowed down or sped up by the gas drag. In this case, inwards of the pressure maximum (and outwards from the pressure minimum at $\sim$\,0.2\,AU) the gas drag acts outwards ($u_{\rm drift}>0$). However, as dust particles are small inside the pressure maximum ($\textrm{St}$\,$\sim$\,$4\times 10^{-4}$ even for the large particles) and their size further decreases inwards, the dust advection velocity is outwards only in a very narrow region. Consequently, after accounting for the diffusivity (i.e., the radial turbulent mixing of dust, which limits the radial gradient of the dust-to-gas ratio), the mass build-up inside the pressure maximum is moderate compared to the rest of the inner disc.

The dust advection velocity used in our method is a mass-weighted average of the velocity of the monomer-sized particles and the large particles (of size $a_{\rm frag}$, Fig. \ref{fig:dust_size}). The monomer-sized particles are just advected by the gas through the pressure maximum, but it can be shown that an individual large particle will also not be trapped. This is due to dust particles being in the fragmentation limit, in which the particles are fragmented faster than they drift. The drift timescale for the large particles inside the pressure trap (the region inwards of the pressure maximum where their advection velocity is outwards) can be estimated by
\begin{equation}
	t_{\rm drift} \approx \frac{1}{2} \frac{d_{\rm trap}}{u_{\rm 1}(r_{P_{\rm max}})} \,,
	\label{eq:drift_timescale}
\end{equation}
where $d_{\rm trap} \approx 0.06$\,AU is the radial width of the trap, and the particle velocity is $u_{\rm 1} \approx 2$\,cm\,s$^{-1}$ (see eq. \ref{eq:advection_velocity}) is evaluated just inwards of the pressure maximum. The velocity $u_{\rm 1}$ decreases inwards, and so this estimate, $t_{\rm drift} \approx 7400$\,yr, is a lower limit. The collisional (i.e., fragmentation) timescale for the large particles, $t_{\rm col} = (n \sigma \Delta v)^{-1}$, is much shorter in comparison. Here $n=f_m \rho_{\rm d} / m_{\rm d}$ is the number density of large particles, $f_m$ is the mass fraction of the large particles \citep[$f_m=0.75$ in the fragmentation limit,][]{Birnstiel2012}, $\rho_{\rm d} \approx \Sigma_{\rm d} \Omega / (\sqrt{2 \pi} c_{\rm s})$ is the midplane mass density of particles, $m_{\rm d}$ is mass of a single particle, $\sigma$ is the collisional cross section and $\Delta v \approx \sqrt{3 \alpha \textrm{St}} c_{\rm s}$ is the typical relative velocity between the particles due to turbulence \citep{OrmelCuzzi2007}. This yields the collisional timescale of:
\begin{equation}
\begin{split}
	t_{\rm col} & = \sqrt{\frac{8}{27\pi}} \frac{\Sigma_{\rm g}}{f_m \Sigma_{\rm d}} \sqrt{\frac{\textrm{St}}{\alpha}} \frac{1}{\Omega} \\
    & \approx 4 
    \left(\frac{f_m}{0.75} \right)^{-1}
    \left(\frac{\Sigma_{\rm d}}{0.01 \Sigma_{\rm g}} \right)^{-1} 
    \left(\frac{\textrm{St}}{10^{-4}} \right)^{1/2} 
    \left(\frac{\alpha}{10^{-4}} \right)^{-1/2} \\
    & \times \left(\frac{\Omega}{10\textrm{\,yr}^{-1}} \right)^{-1} 
    \textrm{yr}
\end{split}
\label{eq:collisional_timescale}
\end{equation}
where we have expressed the particle size $a_{\rm frag}$ in terms of the Stokes number St. At the pressure maximum $\alpha = 10^{-4}$ and so $t_{\rm col} \approx 0.08 \Sigma_{\rm g}/\Sigma_{\rm d}$\,yr\,$\ll t_{\rm drift}$ for dust-to-gas ratios $\Sigma_{\rm d}/\Sigma_{\rm g} \gtrsim 0.01$. Thus, instead of becoming trapped in the pressure maximum, dust particles fragment and flow inwards.

In the innermost disc turbulent fragmentation yields monomer-sized particles that are entrained with the gas ($\bar{u}$\,$\sim$\,$u_{\rm gas}$). And so the radial drift inwards is slowed by particles becoming well coupled to the gas. This is also why, in steady state, the dust-to-gas ratio is roughly constant inwards of the pressure maximum.

Finally, despite the pressure maximum not trapping the inflowing particles, the dust-to-gas ratio is enhanced. Because the dust moves with the gas in the innermost disc, the steady state dust-to-gas ratio there is directly proportional to the ratio of dust-to-gas accretion rates at the outer boundary, $\dot{M}_{\rm d}/\dot{M}_{\rm g}$. Now, the initial dust-to-gas ratio everywhere is $\Sigma_{\rm d}/\Sigma_{\rm g} = 10^{-2}$. Preserving this ratio over time in the inner disc would thus require $\dot{M}_{\rm d}/\dot{M}_{\rm g} = 10^{-2}$ at the outer boundary. However, the growth of dust grains in the outer disc, and the attendant inward radial drift of grains there, means that $\dot{M}_{\rm d}/\dot{M}_{\rm g} > 10^{-2}$ at the outer boundary (i.e., dust accretes inwards preferentially compared to gas). In this case, the inner disc in steady-state will also have $\Sigma_{\rm d}/\Sigma_{\rm g} > 10^{-2}$. In other words, as Fig. \ref{fig:dust_to_gas_ratio} shows, radial drift of grains from the outer disc leads to an enrichment of solids in the inner disc.

What level of the enrichment is attainable depends on the ratio of dust and gas accretion rates $\dot{M}_{\rm d}/\dot{M}_{\rm g}$, i.e. how quickly the grains drift from the outer disc relative to gas accretion. Since the grain growth in the outer disc is limited by radial drift rather than fragmentation \citep{Birnstiel2012}, high grain drift rates are possible. For example, assuming that $\Sigma_{\rm d}/\Sigma_{\rm g}=10^{-2}$ in the outer disc, achieving $\dot{M}_{\rm d}/\dot{M}_{\rm g}=1$ requires the radial drift velocity of grains ($\approx 2 \textrm{ St } u_{\rm drift}$) to reach $10^2 u_{\rm gas}$. For the standard $\alpha$-disc model and $\alpha=10^{-4}$ (extrapolation of the disc model shown in Fig. \ref{fig:gas_structure}) this is satisfied if grains grow to St\,$\sim$\,$10^{-2}$. This corresponds to a particle size an order of magnitude below the radial drift limit ($a_{\rm drift}$) throughout the outer disc. Therefore, the grains easily grow large enough to achieve accretion rates of $\dot{M}_{\rm d}/\dot{M}_{\rm g} \gtrsim 1$. These grains in the outer disc will contain ices which will evaporate as grains drift across the ice lines, towards the inner disc. However, even in the outermost disc silicates account for a considerable portion of the total solid mass \citep[e.g., adopting the abundances of oxygen and carbon in their main molecular carriers from][23\,\% of the total oxygen and carbon mass in the outer disc is in silicates and other refractories]{Oberg2011}. Hence, Fig. \ref{fig:dust_to_gas_ratio} features dust accretion rates up to $\dot{M}_{\rm d}/\dot{M}_{\rm g}=1$ (which, as argued above, are likely), in which case the dust-to-gas ratio in the inner disc also approaches unity.

\subsection{Implications for planetesimal formation} \label{sec:planetesimals}
The above results show that the MRI yields a dust-enhanced inner disc, although at the expense of the dust particle size. Furthermore, as there is no trap for the dust particles, the accumulation of dust is limited by the dust inflow rate from the outer disc, and does not increase indefinitely. In this section we explore if further concentration of particles via the streaming instability \citep[SI;][]{Youdin2005} and subsequent gravitational collapse into planetesimals could be the next step towards forming the close-in super-Earths and mini-Neptunes. 

The SI can greatly concentrate dust particles if the ratio of dust-to-gas bulk densities is $\rho_{\rm d}/\rho_{\rm g}$\,$\gtrsim$\,1 \citep{Youdin2005, Johansen2007}. This is most likely to be attained in the disc midplane, as dust particles gravitationally settle. The settling is balanced by turbulent stirring. One source of turbulence is the MRI. To reach $\rho_{\rm d}/\rho_{\rm g} \geq 1$ in the midplane in the presence of such turbulence, the dust-to-gas surface density ratio $\Sigma_{\rm d}/\Sigma_{\rm g}$ needs to be greater than or equal to $Z_{\rm cr1} = \sqrt{\alpha/(\rm{St}+\alpha)}$ \citep{Carrera2017}.

Even in discs that are weakly turbulent or completely laminar, as dust settles the dust-gas interaction leads to turbulence (self-stirring) which can prevent clumping by the SI. In this case, SI can only successfully concentrate dust particles if the dust-to-gas ratio $\Sigma_{\rm d}/\Sigma_{\rm g}$ is greater than a critical value $Z_{\rm cr2}$ that depends on the particle Stokes number \citep{Johansen2009, Carrera2015}. For St\,$<0.1$ (relevant to our simulations) this critical value has been most recently revised by \citet{Yang2017}, who find 
\begin{equation}
	\log Z_{\rm cr2} = 0.1 \log^2 \rm{St} + 0.2 \log \rm{St} - 1.76 \,.
\label{eq:SI}
\end{equation}
Small dust grains that are entrained with the gas do not participate in the SI. Hence, to compare our results against the SI criteria, we use the dust-to-gas ratio $f_{\rm m} \Sigma_{\rm d}/\Sigma_{\rm g}$ of the large grain population only \citep[where $f_{\rm m}=0.75$ is the mass fraction of large particles;][]{Birnstiel2012}.

\begin{figure}
\centering
\includegraphics[width=\columnwidth, trim={0 1.3cm 0 0}, clip]{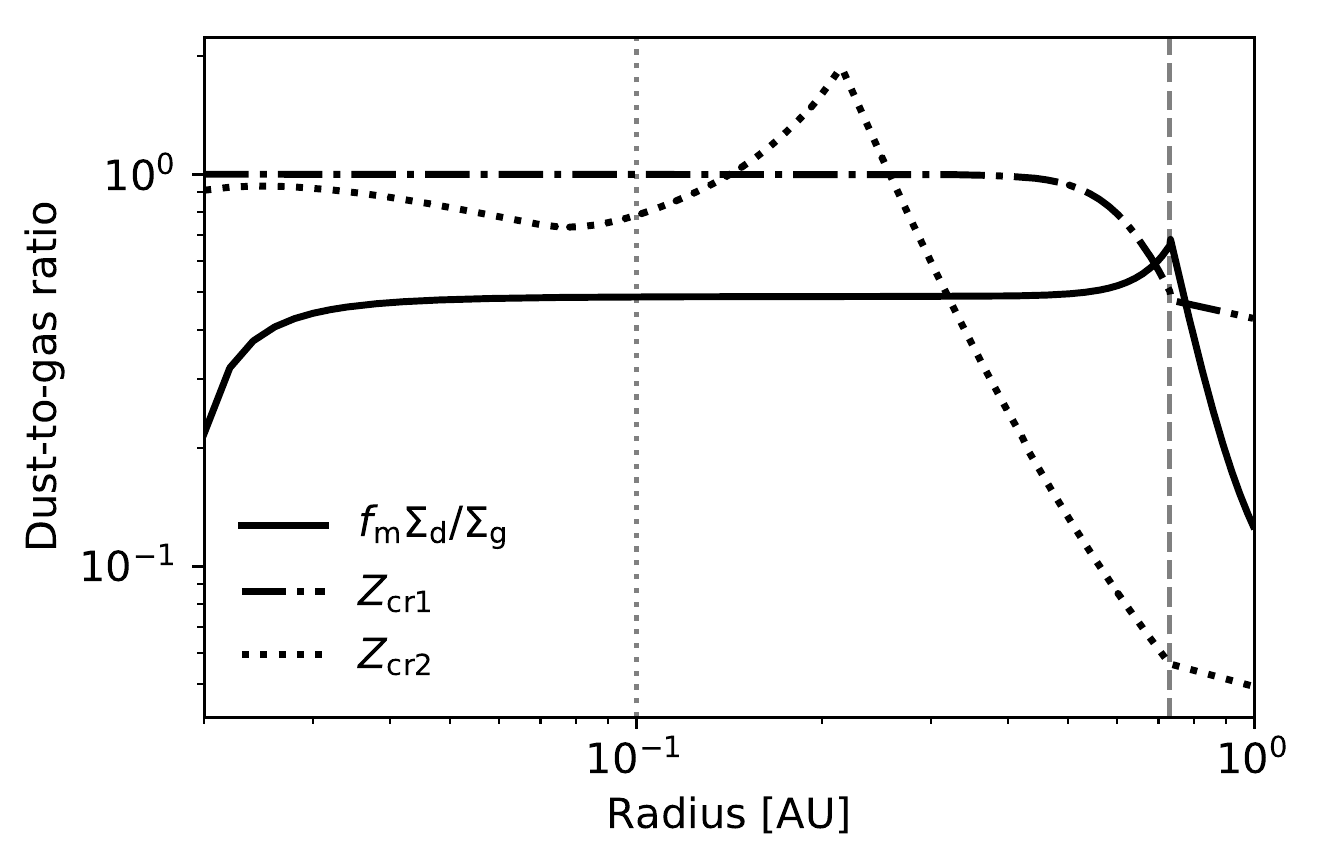}
\includegraphics[width=\columnwidth, trim={0 0 0 0.2cm}, clip]{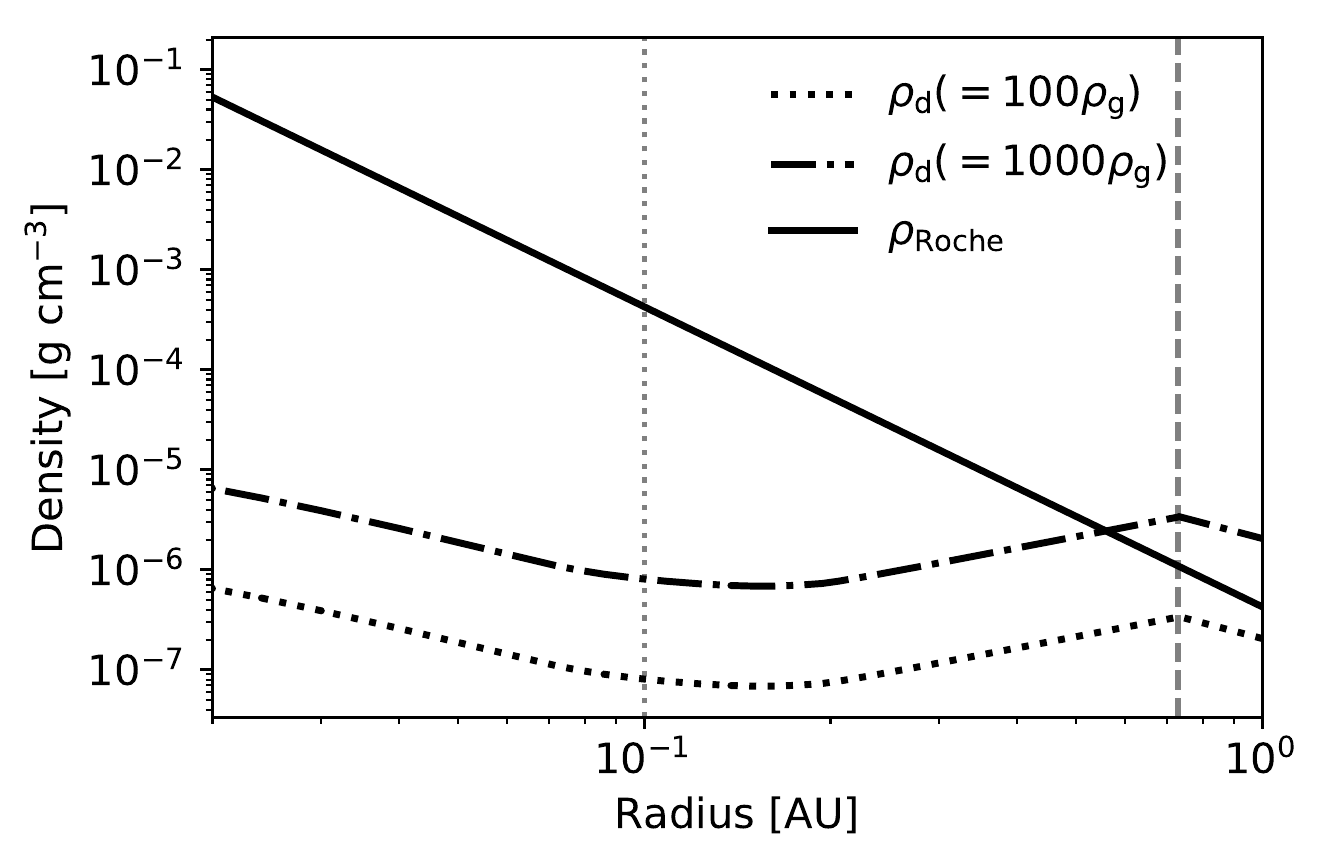}
\caption{\label{fig:SI}
\textbf{Top panel:} Dust-to-gas ratio of large dust grains $f_{\rm m} \Sigma_{\rm d}/\Sigma_{\rm g}$ as a function of radius (when $\dot{M}_{\rm d}/\dot{M}_{\rm g}=1$ at the outer boundary; $f_{\rm m}=0.75$), compared to the two criteria for the onset of the streaming instabilities. \newline
\textbf{Bottom panel:} Expected peak local dust densities $\rho_{\rm d}$ if streaming instabilities successfully concentrate particles in the disc midplane as functions of radius, compared to the Roche density $\rho_{\rm Roche}$.
In both panels the vertical dashed line indicates location of the local pressure maximum, and the dotted line indicates the location of the dust sublimation line.}
\end{figure}

In the top panel of Fig. \ref{fig:SI} we compare this dust-to-gas ratio, for the outer boundary condition $\dot{M}_{\rm d}/\dot{M}_{\rm g}=1$, with the above two criteria for the onset of the SI. We find that in the inner disc, turbulence due to the MRI is generally more prohibitive to dust settling than the turbulence due to dust-gas interactions. Both conditions are fulfilled only near the pressure (and density) maximum.

Provided that the SI successfully concentrates dust particles in the disc midplane, the dust bulk density there may reach up to 100\,--\,1000 times the local gas density \citep{Johansen2007}. Gravitational collapse of such particle concentrations will occur if the dust density exceeds the local Roche density (below which the star can tidally disrupt the fragment), $\rho_{\rm Roche}=9 M_{\ast}/ 4\pi r^3$. Comparing $\rho_{\rm Roche}$ with the midplane gas densities in our steady-state MRI disc (Fig. \ref{fig:SI}, bottom panel), we find that the condition for gravitational collapse is only satisfied near the pressure (and density) maximum, or at larger separations. Importantly, the bottom panel of Fig. \ref{fig:SI} shows that the possibility of gravitational collapse of solids in the inner disc is severely limited by the steep gradient of the Roche density.

Overall, given the stellar and disc parameters used here, we find that the SI and the gravitational collapse pathway to planetesimals is viable in the inner disc only in a very narrow region near the pressure maximum. For gas accretion rates larger than the one used here \citep[$\dot{M}_{\rm g} > 10^{-8}$\,M$_\odot$\,yr$^{-1}$, that could be expected in the early phase of disc evolution, e.g.][]{Manara2012} this conclusion will hold, while for sufficiently smaller accretion rates planetesimals would not form in this way anywhere near or inwards of the pressure maximum. Firstly, in both cases dust evolution is expected to yield roughly the same steady-state dust-to-gas ratios given the same ratio of dust and gas accretion rates at the outer boundary. To confirm this conclusion, we repeat the dust evolution calculations for the gas accretion rate of $\dot{M}_{\rm g}=10^{-9}$\,M$_\odot$\,yr$^{-1}$, obtaining results very similar to those above. In addition to similar steady-state dust-to-gas ratios, the particle Stokes number sharply drops inwards of the pressure maximum, and the SI is similarly triggered only around the pressure maximum. This is because the slope of the increase in $\alpha$ inwards of the pressure maximum is roughly the same for different $\dot{M}_{\rm g}$. Hence the SI criteria is expected to be fulfilled only near the pressure maximum for higher gas accretion rates as well.

Secondly, even if SI is successfully triggered, to form planetesimals the peak dust density needs to be above the Roche density. The peak dust density scales with the midplane gas density at the pressure maximum, so it scales with the gas accretion rate approximately as $\dot{M}_{\rm g}^{-1/2}$, and with the radial location of the pressure (and density) maximum as $r_{P_{\rm max}}^{-1}$. Since the peak dust density is larger than the Roche density at the pressure maximum for $\dot{M}_{\rm g}=10^{-8}$\,M$_\odot$\,yr$^{-1}$, and the Roche density $\rho_{\rm Roche} \propto r^{-3}$, for $\dot{M}_{\rm g}>10^{-8}$\,M$_\odot$\,yr$^{-1}$ the gravitational collapse criterion will also be fulfilled near the pressure maximum, while for a sufficiently smaller accretion rate (including $\dot{M}_{\rm g}=10^{-9}$\,M$_\odot$\,yr$^{-1}$) peak dust density will be too low.

Moreover, it is important to note that, in order for the SI to operate, there must be a relative azimuthal velocity between the dust and gas \citep[e.g.][]{Squire2018}, in addition to the density criteria invoked above. Such a relative velocity necessarily disappears at the pressure maximum itself, further constricting the region where the SI is viable in our disc.

However, we reiterate that our calculations do not include the effects of dust on the gas dynamics and on the MRI. The latter effect in particular may relax some of the constraints on the SI, and we discuss this possibility in section \ref{sec:discussion}. Overall, if the close-in sub-Neptunes form \textit{in situ}, in an MRI-accreting inner disc \citep[as suggested by][]{Chatterjee2014}, any theory needs to explain how the dust grains can grow larger to either start the SI in a wider vicinity of the pressure maximum, or for the pressure maximum to become an efficient dust trap that can concentrate grains to densities needed for the gravitational collapse, or find an alternative pathway to planetesimal and core formation.

\section{Accretion and evolution of planetary atmospheres} \label{sec:atmospheres} 

We showed above that planetesimal formation through streaming instabilities and gravitational collapse is challenging despite the dust enhancement in the inner disc. Although these conclusions could change (see section \ref{sec:discussion}), it is not presently clear how exactly planetesimals or cores would arise in the inner disc, and we are unable to predict properties of solid cores formed in situ. Nevertheless, orbital distances, radii and masses of many close-in planets have been well determined observationally. Furthermore, their radius distribution has been shown to be consistent with Earth-like composition and thus formation inside the ice line \citep{Owen2017}, possibly in the inner disc. Thus, in this section we use observational results to perform a separate test of in situ formation by considering accretion of planetary atmospheres in the inner disc and their subsequent evolution.

To follow the accretion of planetary atmospheres, we assume that solid super-Earth sized cores are present in the inner regions of our gas-poor inner disc structure taken from \citet{Mohanty2017arxiv}. Furthermore, following the arguments of \citet{Lee2016}, we ignore additional heating arising from further accretion of solids, as any amount of accretion providing significant heating typically results in the core rapidly reaching run-away masses.

After disc dispersal (which we do not model), we also account for atmospheric mass lost due to photoevaporation. This is an important addition, as several theoretical \citep[e.g.][]{Owen2013,Lopez2013} and observational \citep[e.g.][]{Lundkvist2016,Fulton2017,Fulton2018} studies have shown that photoevaporation significantly sculpts the exoplanet population after formation.

\subsection{Methods} \label{sec:atmospheres_methods}
\subsubsection{Accretion of planetary atmospheres}
We assume that solid super-Earth-sized planet cores accrete their gaseous envelopes in a gas-poor inner disc that is viscously accreting due to the MRI; this disc structure does not evolve in time. For a quasi-steady state envelope %placed in a gas-poor inner disc ($\Sigma_g$\,$\sim$\,$1000$\,g\,cm$^{-2}$), e.g. at an orbital distance of 0.3\,AU in the MRI-accreting inner disc, 
the accreted envelope mass fraction after time $t$ can be estimated by the scaling relations from \citet{Lee2015} \citep[with an additional factor that accounts for varying gas surface density from][]{Lee2017, FungLee2018}:
\begin{eqnarray}
	X(t) &=& 0.07 \left(\frac{t}{1\textrm{\,Myr}} \right)^{0.4} 
    \left(\frac{0.02}{Z} \right)^{0.4} 
    \left(\frac{\mu}{2.37} \right)^{3.4}
    \nonumber\\ 
    &&\times
    \left(\frac{M_{\rm core}}{5 \textrm{M}_\oplus} \right)^{1.7} 
    \left ( \frac{f_\Sigma}{0.1} \right)^{0.12}
    \label{eq:dusty}
\end{eqnarray}
for dusty atmospheres, and
\begin{eqnarray}
	X(t) &=& 0.18 \left(\frac{t}{1\textrm{\,Myr}} \right)^{0.4} 
    \left(\frac{0.02}{Z} \right)^{0.4} 
    \left(\frac{\mu}{2.37} \right)^{3.3}\nonumber\\ 
    &&\times \left(\frac{M_{\rm core}}{5 \textrm{M}_\oplus} \right)^{1.6} 
    \left(\frac{1600\textrm{\,K}}{T_{\rm rcb}} \right)^{1.9}
    \left ( \frac{f_\Sigma}{0.1} \right)^{0.12}
    \label{eq:dust-free}
\end{eqnarray}
for dust-free atmospheres. Here $Z$ is the metallicity of the atmosphere; $\mu=1/(0.5 W + 0.25 Y + 0.06 Z)$ is the mean molecular weight, with $W=(1-Z)/1.4$, $Y=0.4(1-Z)/1.4$; $T_{\rm rcb}$ is the temperature at the radiative-convective boundary inside the atmosphere; $f_\Sigma = \Sigma_{\rm g}/\Sigma_{\rm MMSN}$ is the ratio of the gas surface density ($\Sigma_{\rm g}$ from our inner disc model, Fig. \ref{fig:gas_structure}) and the gas surface density profile of the minimum mass solar nebula \citep[$\Sigma_{\rm MMSN} = 1700 (d/1\textrm{\,AU})^{-3/2}$\,g\,cm$^{-2}$, where $d$ is the orbital radius,][]{Hayashi1981}. %\citet{Lee2015} include an additional fitting parameter that varies with distance and surface density; however, it is typically of order unity and varies only weakly with both parameters, so we neglect this variation here. 
Furthermore, we have assumed the gas adiabatic index is $\gamma=1.2$, and that in dusty atmospheres $T_{\rm rcb}=2500$\,K which arises from the disassociation of Hydrogen (see \citet{Lee2015}, their section 2.1).

We use the expressions (\ref{eq:dusty},\ref{eq:dust-free}) to calculate how much gas a planet accretes in 1\,Myr as a function of core mass, for various metallicities $Z$ and gas surface density factors $f_\Sigma$ in the case of dusty atmospheres (with $T_{\rm rcb}$ = 2500\,K), and various $Z$, $f_\Sigma$ and $T_{\rm rcb}$ in the case of dust-free atmospheres. 

\subsubsection{Photoevaporation of planetary atmospheres}
These accreted atmospheres are then subjected to photoevaporation following disc dispersal. We use a simplified estimate of the photoevaporative mass loss. First, for a given planet core mass $M_{\rm core}$ and (accreted) envelope mass fraction $X$ we find the photospheric radius of the planet $R_{\rm p}$. For this we use a simple model of an atmosphere at hydrostatic equilibrium \citep{Owen2017}, in which the solid core is surrounded by an adiabatic convective envelope, topped by an isothermal radiative photosphere.% The temperature of the photosphere is set by stellar insolation, and we adopt the same opacity law as \citet{Owen2017}, corresponding to a dust-free H/He envelope of solar metallicity.
%The atmosphere is gravitationally contracting and cooling on {\jo its} Kelvin-Helmholtz timescale $\tau_{\rm KH}$. Following \citet{Owen2017} we set $\tau_{\rm KH} = 100$\,Myr for the first 100\,Myr, after which $\tau_{\rm KH}$ equals planet age. We do not, however, account for the mass loss due to the ``boil-off'', and we discuss how it may affect our results in section \ref{sec:discussion}.

Next, the mass-loss timescale due to high-energy stellar irradiation is \citep{Owen2017}
\begin{equation}
	t_{\dot{X}} = \frac{4\pi d^2 G M_{\rm core}^2 X(1+X)}{\eta \pi L_{\rm HE}} \frac{1}{R_{\rm p}^3} \,,
	\label{eq:tloss}
\end{equation}
where $d$ is the orbital radius of the planet and $L_{\rm HE}$ is the stellar high-energy flux. We consider a Sun-like star, as in our disc model above.

To determine the final envelope mass fractions, we do not explicitly evolve the atmospheres in time. Instead, we use the fact that most of the mass loss happens during the first $\sim$\,100\,Myr after disc dispersal, since during this period the stellar high-energy flux $L_{\rm HE}$ is saturated ($L_{\rm HE}=L_{\rm sat} \sim 10^{-3.5} L_\odot$ for a Sun-like star) while after this time it quickly decays \citep{Jackson2012,Tu2015}. 

Thus, if a planet's mass-loss timescale $t_{\dot{X}}$ is initially (i.e., at the time of disc dispersal) longer than 100\,Myr, the planet will not suffer significant mass loss. Here we assume that such a planet remains unchanged by the photoevaporation. 

On the other hand, a planet with initial $t_{\dot{X}}<100$\,Myr will lose mass. Now, for a given core mass $M_{\rm core}$ and orbital distance $d$, the mass-loss timescale as a function of the envelope mass fraction, $t_{\dot{X}}(X)$, peaks at $X \equiv X_{\rm peak}$ of a few percent, decreasing for both smaller and larger $X$ \citep{Owen2017}. Thus, for a planet with an initially small accreted envelope mass fraction ($X < X_{\rm peak}$), the mass loss further shortens the loss timescale. Hence, if such a planet's initial mass-loss timescale is less than 100\,Myr it is subject to runaway mass loss, and we assume it is completely stripped of its atmosphere. Conversely, the mass-loss timescale of a large atmosphere ($X > X_{\rm peak}$) increases as it loses mass, tending towards the peak value of $t_{\dot{X}}(X_{\rm peak})$. If $M_{\rm core}$ and $d$ are such that the latter peak timescale is $\geq$100\,Myr, we assume that a planet with such a large initial atmosphere will stall at an envelope mass fraction $X$ corresponding to a loss timescale of $t_{\dot{X}}$ = 100\,Myr. However, if the peak timescale is $<$100\,Myr, the mass-loss timescale can increase to this peak value and then descend into the runaway regime on the other side while the stellar activity is still saturated; thus, such a planet will lose its entire atmosphere regardless of the accreted $X$. This simple prescription adequately captures the basic physics of atmospheric photoevaporation \citep[see far left panel of Fig.\,6 in][]{Owen2017}.

\subsection{Results} \label{sec:atmospheres_results}
\subsubsection{Accretion of planetary atmospheres}
\begin{figure}
\centering
\includegraphics[width=\columnwidth]{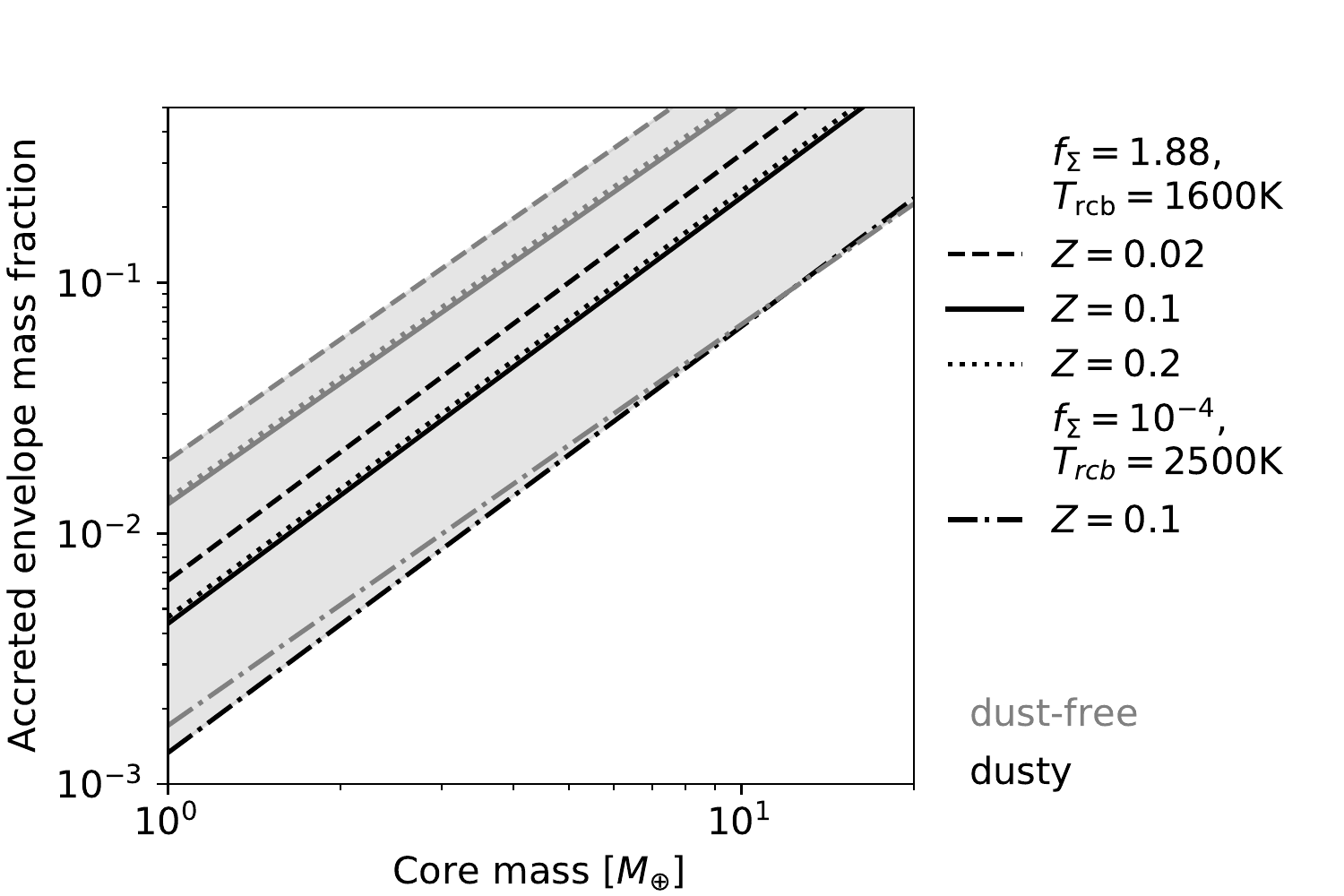}
\caption{\label{fig:accreted_atmospheres}
Envelope mass fraction of atmospheres accreted in 1\,Myr as a function planet core mass for dusty (black lines) and dust-free (grey lines) atmospheres and a variety of metallicities $Z$, gas surface density factors $f_\Sigma$ and (in the case of dust-free atmospheres) radiative-convective boundary temperatures $T_{\rm rcb}$, as indicated in plot legend. Grey region indicates the total range of expected envelope mass fractions (except for those that would reach an envelope mass fraction of $X=0.5$ within 1\,Myr and thereby expected to undergo runaway accretion to form gas giants; these are not shown).}
\end{figure}

Using the scaling relations (\ref{eq:dusty},\ref{eq:dust-free}) we calculate the envelope mass fractions that planetary cores of various masses accrete from the gas disc in 1\,Myr. Results are shown in Fig. \ref{fig:accreted_atmospheres} for both dusty and dust-free atmospheres of various metallicities, ranging from solar ($Z=0.02$) to the metallicity of Neptune's atmosphere \citep[$Z=0.2$; ][]{Karkoschka2011}, % from methane measurements at low altitudes number abundance of methane is 0.04, so mass abundance of methane is ~0.25, out of which 12/16 part is metals i.e. carbon
gas surface density factors $f_\Sigma=10^{-4}$\,--\,1.88, and radiative-convective boundary temperatures $T_{\rm rcb}=1600$\,--\,2500\,K.

The dependence on metallicity $Z$ is non-monotonous, as the accreted envelope mass fraction depends separately on the metallicity and on the mean molecular weight (which is set by the metallicity). The smallest accreted atmospheres, with the rest of the parameters fixed, have $Z$\,$\sim$\,$0.1$.

The radiative-convective boundary temperature $T_{\rm rcb}$ is expected to be roughly constant in dusty atmospheres, so we only explore the effect of this parameter in dust-free atmospheres. In the latter, $T_{\rm rcb}$ is related to the temperature of the environment. Additionally, accretion of both dust-free and dusty atmospheres depends the density of the environment. Here we are interested in atmospheres that are accreted in the inner disc, near or inwards of the pressure maximum. The location of the pressure maximum is determined by the extent of thermal ionization of potassium in our disc model, and so this corresponds to disc temperatures of $T$\,$\gtrsim$\,1000\,K, regardless of the exact disc parameters (e.g. gas accretion rate). For a disc temperature of $T$\,$\sim$\,1000\,K, numerical models of the accreting atmospheres give $T_{\rm rcb}$\,$\sim$\,$1600$\,K \citep{Lee2015}, which thus sets a lower bound on $T_{\rm rcb}$ for our calculations. Moreover, the location of the pressure maximum is also where the gas surface density is highest (see Fig. \ref{fig:gas_structure}). Colder atmospheres in a more dense environment accrete more. So, to show the maximum accreted atmospheres in an MRI-accreting disc, we plot a set of dusty and dust-free atmospheres (of various metallicities $Z$) for the maximum $f_\Sigma=1.88$, and the minimum $T_{\rm rcb}=1600$\,K (the latter refers only to the dust-free atmospheres). Conversely, the maximum temperature at which equation (\ref{eq:dust-free}) is valid \citep[due to the limitations of the opacity tables used by][]{Lee2014} is $T_{\rm rcb}=2500$\,K%and so in Fig. \ref{fig:accreted_atmospheres} we show the dust-free accreted atmospheres corresponding to these two limiting temperatures
\footnote{Note that, assuming $T_{\rm rcb}$ is directly proportional to the disc temperature $T$, and scaling from the numerical models' result that $T_{\rm rcb}$\,$\sim$\,1600\,K corresponds to $T$\,$\sim$\,1000\,K, yields $T$\,$\sim$\,1500\,K for $T_{\rm rcb}$\,$\sim$\,2500\,K. The disc temperature in our model only exceeds 1500\,K at radii $<$0.1\,AU, so a maximum $T_{\rm rcb}$ of 2500\,K is indeed roughly valid over most of our inner disc.}, and the minimum gas surface density in our inner disc model with respect to the minimum mass solar nebula is $f_\Sigma \approx 10^{-4}$ (corresponding to the inner disc edge in Fig. \ref{fig:gas_structure}). Hotter atmospheres in lower-density environments accrete less, and so to show the smallest accreted dust-free atmospheres, we plot the $f_\Sigma=10^{-4}$ atmospheres, with $T_{\rm rcb}=2500$\,K for the dust-free atmospheres, and with metallicity $Z=0.1$ (since, as noted above, $Z$\,$\sim$\,0.1 yields the smallest atmosphere for any given set of other parameters).

Finally, atmospheres that grow above a threshold of $X=0.5$ undergo runaway accretion and end up as gas giants \citep{Rafikov2006}. The scaling relations (\ref{eq:dusty},\ref{eq:dust-free}) are not applicable in this case, neither are we interested in the much rarer close-in Jupiters. Therefore, Fig. \ref{fig:accreted_atmospheres} is cut off at $X=0.5$, and the grey region indicates how small or large super-Earth/mini-Neptune atmospheres may be at the time of disc dispersal. Overall we see that, if the cores are formed 1\,Myr before the dispersal, runaway accretion is avoided for the majority of relevant core masses, but they do accrete significant gaseous envelopes of up to a few\,$\times$\,10\% of core mass.

The envelopes shown in Fig. \ref{fig:accreted_atmospheres} have been calculated assuming that the accretion lasts for 1\,Myr. Since disc lifetimes can be longer \citep{Mamajek2009}, these envelopes could be conservative estimates if planets form sooner than 1\,Myr before disc dispersal. If, for example, the envelopes are accreted for 5\,Myr, the envelope mass will double. We do not expect the results to be very sensitive to the exact disc parameters, as long as the cores accrete their atmospheres in a thermally-ionized MRI-active inner disc. We note, however, that our disc model implies the extent of such inner disc does not encompass all observed sub-Neptunes for all relevant accretion rates; e.g. the gas pressure maximum is at an orbital period longer than 100 days only for gas accretion rates of $\dot{M}_{\rm g}$\,$\gtrsim$\,$3\times 10^{-9}$\,M$_\odot$\,yr$^{-1}$. Thus, a planet with  a longer orbital period might spend at least some time in a colder MRI-dead zone, which we do not take into account.

\subsubsection{Photoevaporation of planetary atmospheres}
\begin{figure}
\centering
\includegraphics[width=\columnwidth]{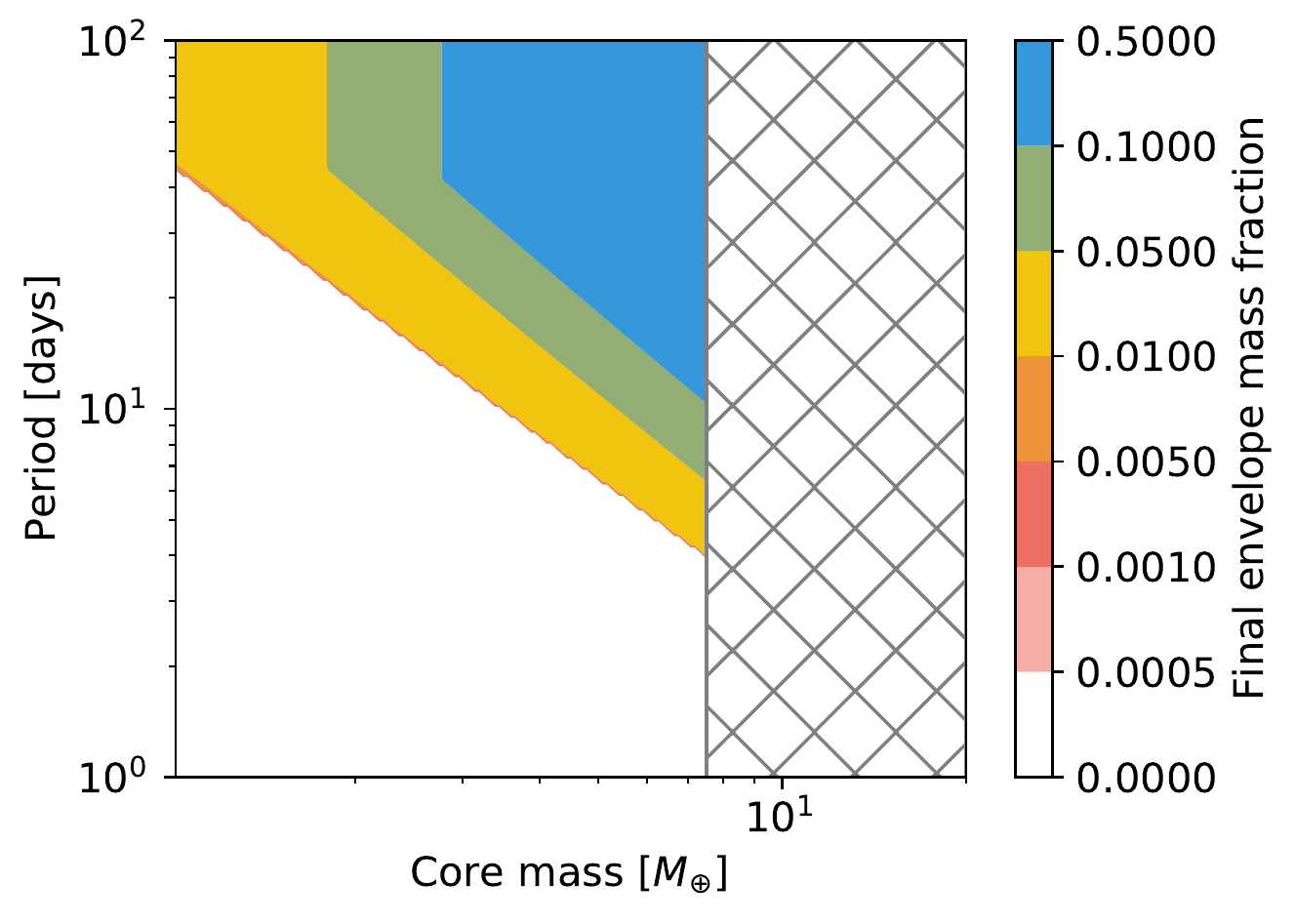}
\includegraphics[width=\columnwidth]{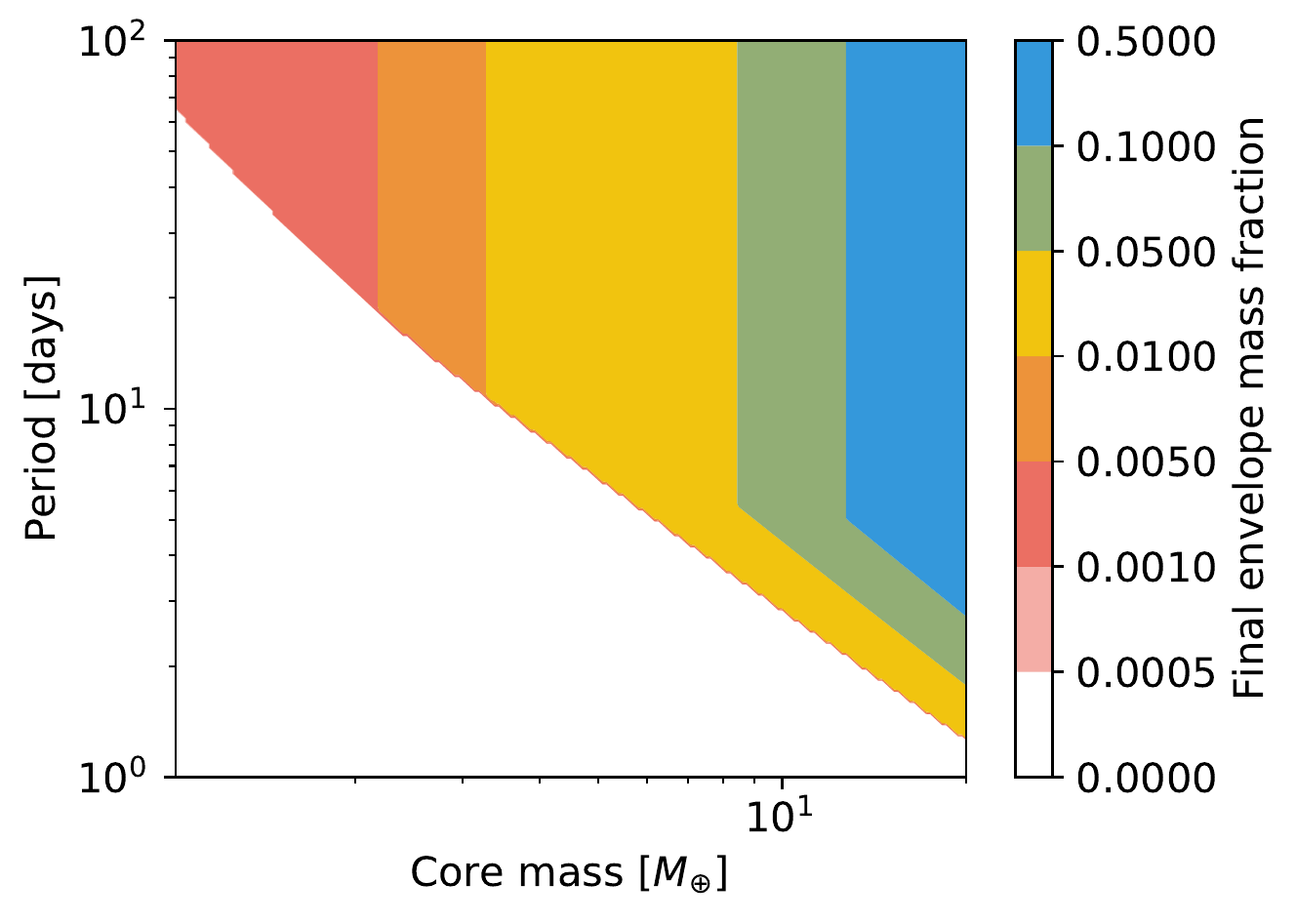}
\caption{\label{fig:evaporated_atmospheres}
Maximum (\textbf{top}) and minimum (\textbf{bottom}) envelope mass fraction of the atmospheres after accounting for photoevaporation, as functions of planet core mass and orbital period. In the top figure, the hatched region indicates the core masses for which the planets would undergo runaway accretion and are thus excluded from here.}
\end{figure}

To further check the consistency of core accretion in the MRI-accreting inner disc with observations, we need to consider whether these accreted atmospheres survive photoevaporation. We calculate the final (remaining) envelope mass fraction of the minimum and maximum possible accreted atmospheres (corresponding respectively to dusty atmospheres with $f_\Sigma=10^{-4}$ and $Z=0.1$, and dust-free atmospheres with $Z=0.02$, $f_\Sigma=1.88$ and $T_{\rm rcb}=1600$\,K) for each core mass and as a function of orbital period. Results are shown in Fig. \ref{fig:evaporated_atmospheres}. In the case of the maximum accreted atmospheres (top panel), the atmospheres would undergo runaway accretion for core masses $\gtrsim$\,8\,M$_\oplus$ (indicated by the hatched region), which are thus excluded here.

The figures show that the orbital period at which the atmosphere can be completely evaporated decreases with increasing core mass, and cores that retain their atmospheres generally evolve towards a 1\% envelope mass fraction as expected from theory. At 100 days the atmospheres are unaffected by photoevaporation, and at periods shorter than 1 day all planets are predicted to end up as bare cores. Massive cores are predicted to keep their 1\,--\,50\% atmospheres at the majority of orbital periods, and planets with Earth-mass cores are safe from complete mass loss at periods larger than 50 days.

Note that here the orbital period determines the level of high-energy flux that planet experiences and planet equilibrium temperature (and thus planet radius), but does not directly reflect variations in temperature and density of the protoplanetary disc inside which the atmospheres were accreted. As discussed above, the effect of the disc temperature on the accreted envelope mass fraction is negligible for dusty atmospheres. For dust-free atmospheres the dependence is monotonous and the extent of the effect is explored by considering the minimum $T_{\rm rcb}$ we expect in the inner disc, and the maximum $T_{\rm rcb}$ for which the scaling relations (\ref{eq:dusty},\ref{eq:dust-free}) are valid. Similarly, the dependence on the disc density is explored by considering the smallest and largest values of the ratio of the MRI-disc model and the minimum mass solar nebula surface densities. Thus, by calculating the effect of photoevaporation on both the minimum and maximum accreted atmospheres shown in Fig. \ref{fig:accreted_atmospheres} for each core mass, we also encompass the possible range of disc densities and temperatures. %The variations in the gas surface density are not included, and we briefly discuss this small correction in section \ref{sec:discussion}.

\subsection{Comparison to observations} \label{sec:atmospheres_comparison}
\begin{figure}
\centering
\includegraphics[width=\columnwidth]{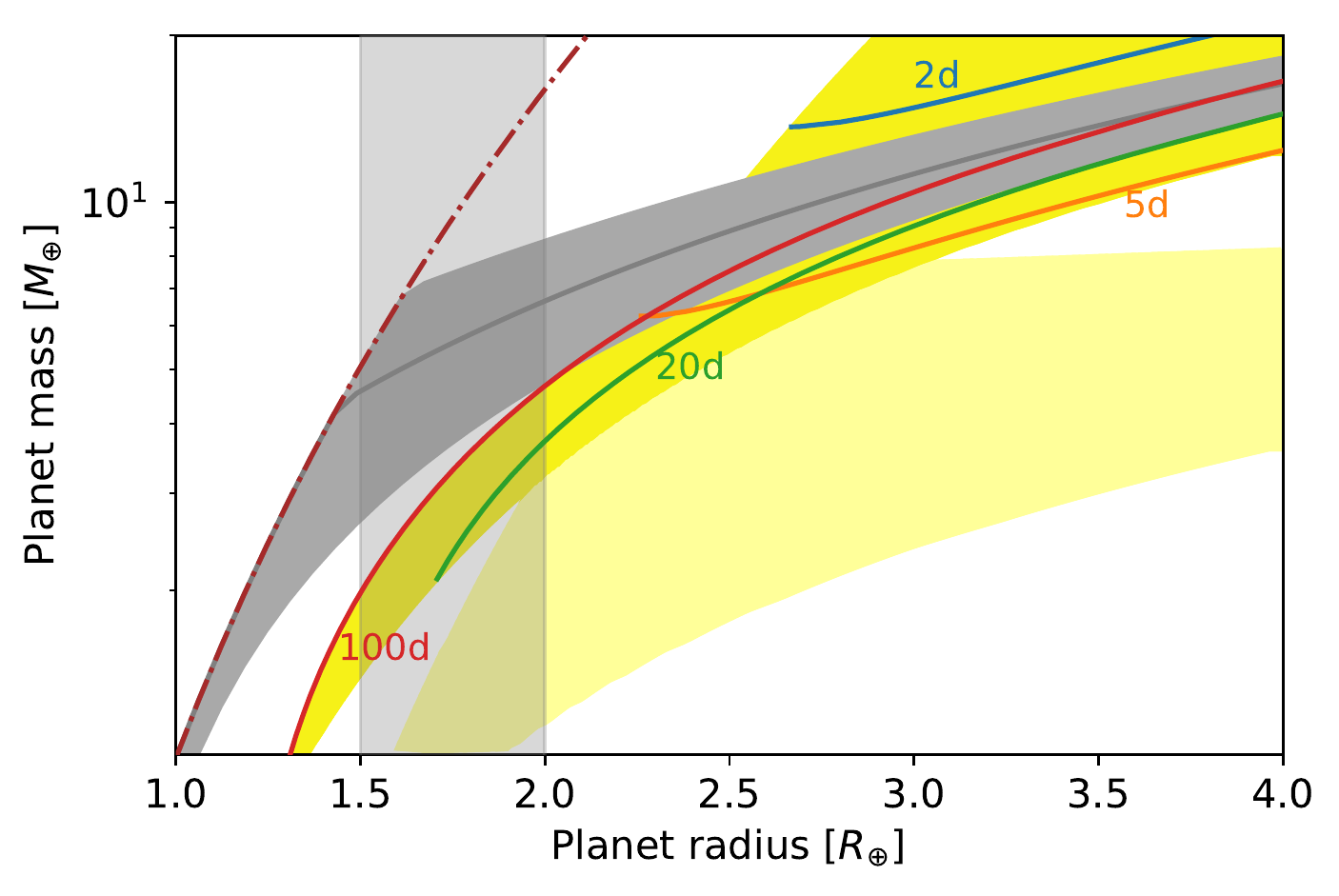}
\caption{\label{fig:MR_relationship}
Mass-radius relationship for sub-Neptune planets: Earth-like composition solid cores (dash-dot-dash line), probabilistic fit to observations mean value (solid grey line) and scatter (solid grey region) \citep{Wolfgang2016}, region of low planet occurrence rates from the observed radius distribution of planets (sheer grey region) \citep{Fulton2017}, and predictions from the minimum and maximum accreted atmospheres and photoevaporation (yellow and light yellow region respectively) with orbital period contours for the minimum accreted atmospheres (solid-line contours).}
\end{figure}

\begin{figure}
\centering
\includegraphics[width=0.95\columnwidth, trim={0 1.3cm 0 0}, clip]{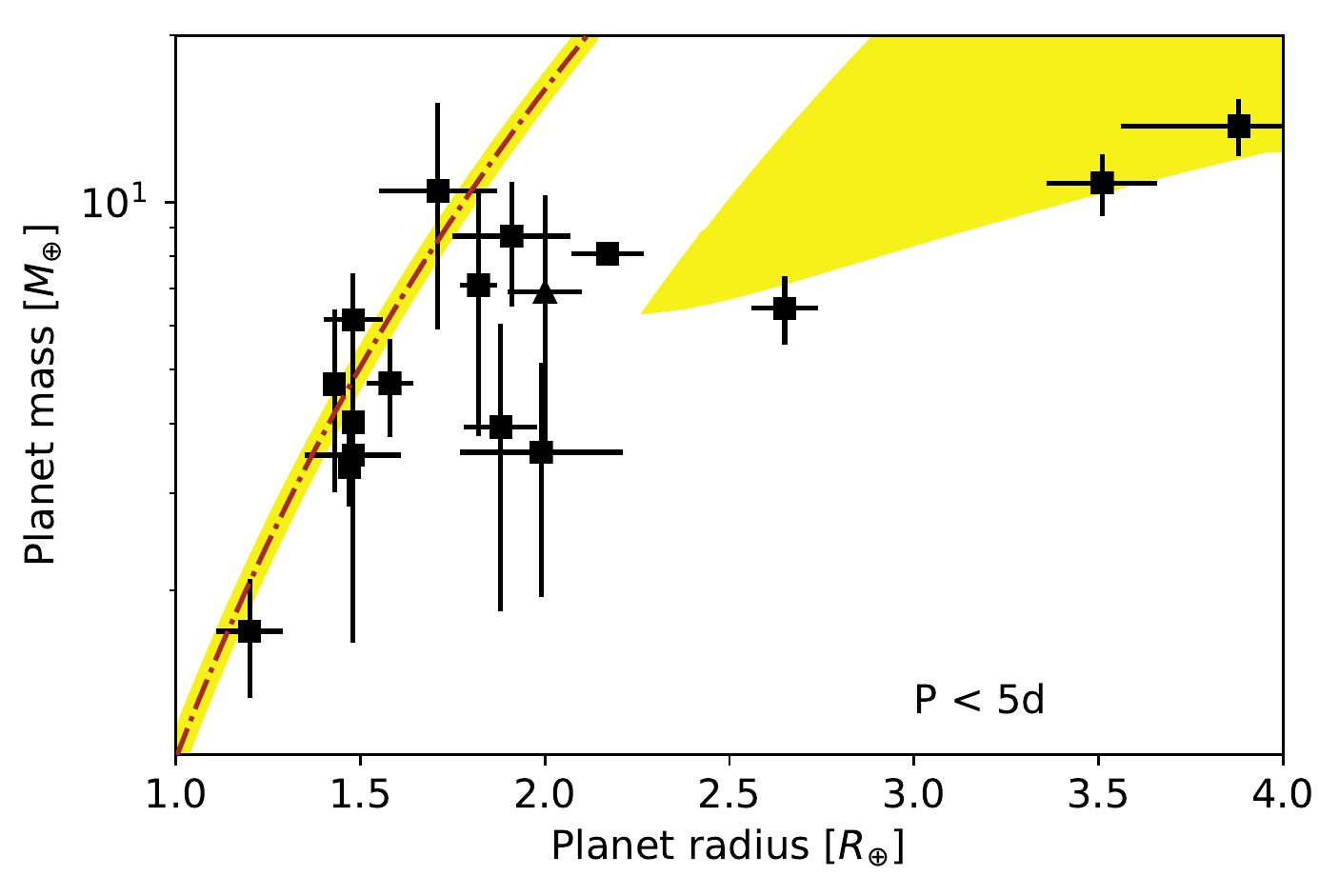}
\includegraphics[width=0.95\columnwidth, trim={0 1.3cm 0 0}, clip]{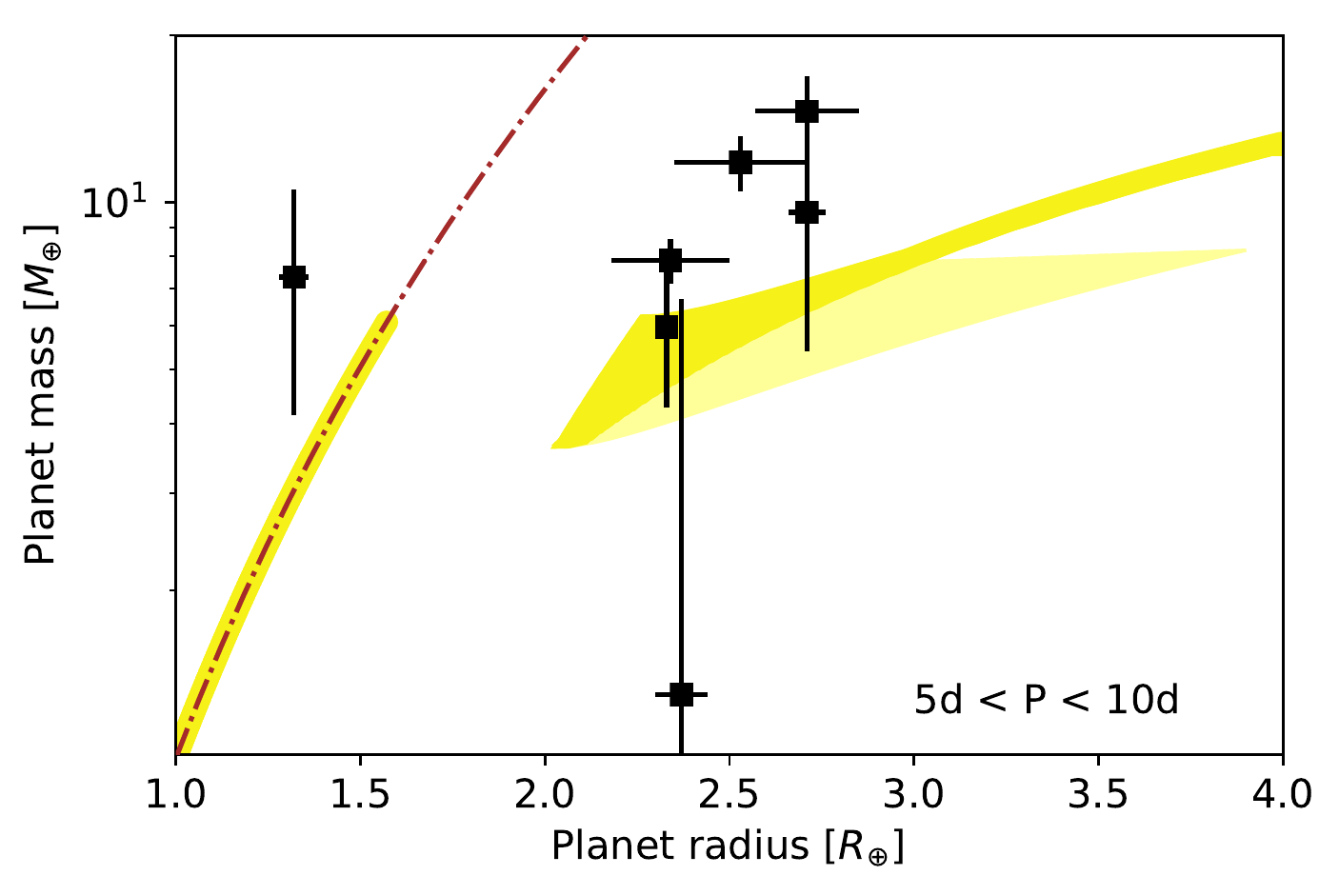}
\includegraphics[width=0.95\columnwidth, trim={0 1.3cm 0 0}, clip]{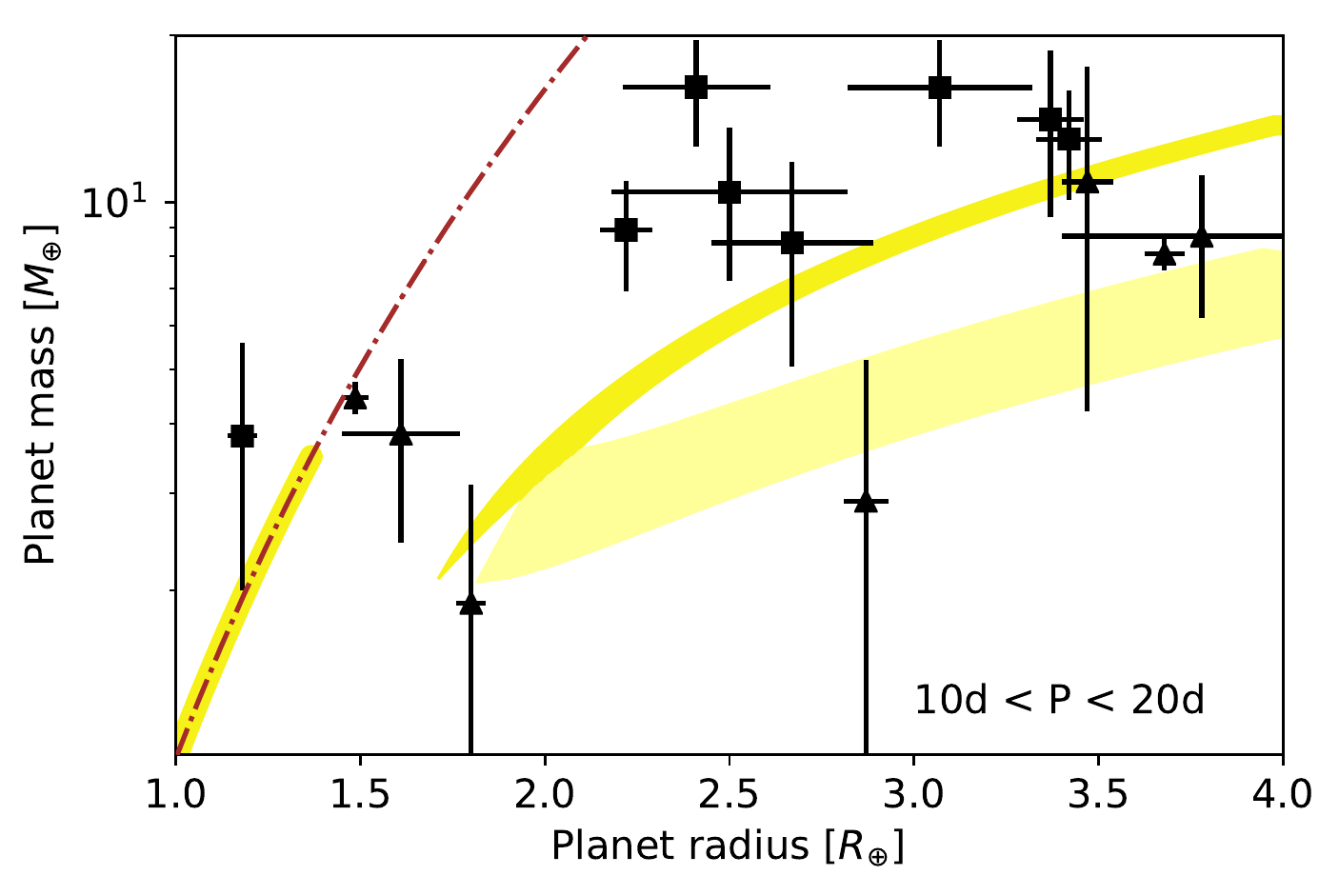}
\includegraphics[width=0.95\columnwidth]{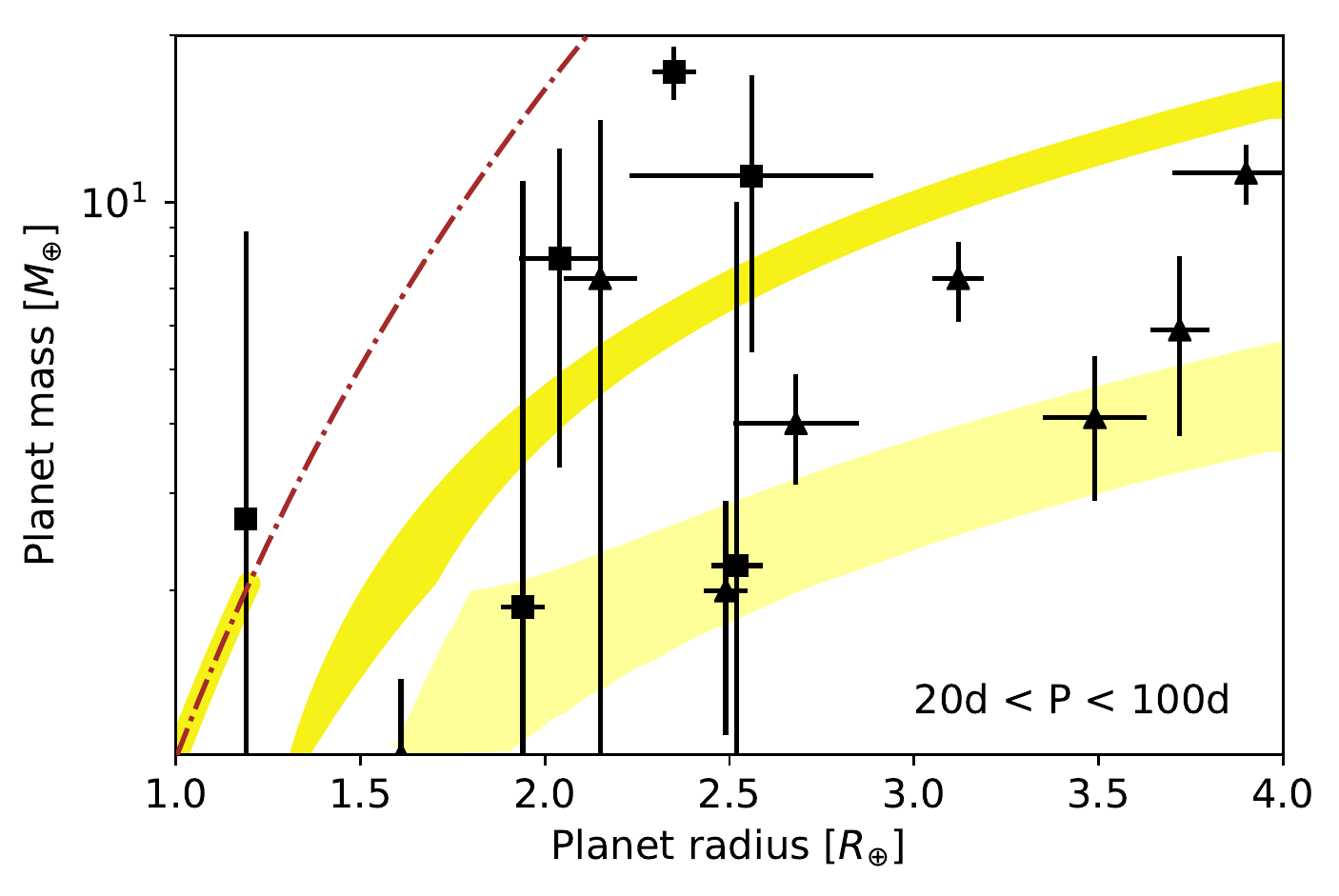}
\caption{\label{fig:MR_ind_planets}
Sub-Neptune planets with measured masses and radii (square markers if the mass was determined using the radial velocity method, and triangles if the mass was determined using transit-timing variations), with uncertainties as listed in \citet{Wolfgang2016}, Earth-like composition solid cores (dash-dot-dash line), and predictions from the minimum and maximum atmosphere models (yellow and light yellow region respectively), in period bins as indicated in plot labels. This figure indicates that while the period range at which planets can be stripped by photoevaporation is consistent with the data, our planets typically have larger H/He envelopes that expected.}
\end{figure}

Overall, Fig. \ref{fig:evaporated_atmospheres} shows that the envelopes formed in a gas-poor inner disc due to the MRI survive photoevaporation for a large range of orbital periods, and the low gas surface densities are not a hindrance to the formation of mini-Neptunes. On the contrary, the final envelope mass fractions of the planets that do keep their atmospheres are typically overestimated. The planets with core mass larger than 2\,M$_\oplus$ are predicted to either have a $>1\%$ atmosphere or to be completely evaporated. On the other hand, from the observations, the typical envelope mass fraction of mini-Neptunes that hold onto their atmospheres is 1\% \citep{Wolfgang2015}. To look into this further, we compare the predictions of our calculations against the observed mass-radius relationship for sub-Neptune planets in Fig. \ref{fig:MR_relationship}, and against measured masses and radii of individual sub-Neptune planets in Fig. \ref{fig:MR_ind_planets}.

For the observations in Fig. \ref{fig:MR_relationship}, we show the probabilistic best-fitting mass-radius relationship of \citet{Wolfgang2016}: a power law $M/\textrm{M}_\oplus = 2.7 (R/\textrm{R}_\oplus)^{1.3}$ (indicated by the solid grey line) with a standard deviation of $\pm 1.9$\,M$_\oplus$ due to an intrinsic scatter in planet mass (the grey region), and an upper limit constraint on the planet density corresponding to a mass-radius relationship for solid cores of Earth-like composition $M/\textrm{M}_\oplus = (R/\textrm{R}_\oplus)^4$ \citep[dash-dot-dash line;][]{Valencia2010}. Additionally, the above mass-radius relationship does not capture a significant feature of the observed radius distribution of sub-Neptunes, a decrease in occurrence rates of planets with radii of 1.5\,--\,2\,R$_\oplus$ \citep[indicated here by the sheer grey region;][]{Fulton2017}.

To show the predictions of our atmospheric calculations in the mass-radius plane, we take the calculated envelope mass fraction as a function of core mass and period and re-calculate the planet radius at the planet age of 5\,Gyr as a function of core mass and period, using the same simple atmospheric evolution model of \citet{Owen2017}. We show the results for the minimum and maximum accreted atmospheres (Fig. \ref{fig:evaporated_atmospheres}, and excluding the completely evaporated planets) in Fig. \ref{fig:MR_relationship} (the yellow and the light yellow region respectively). Note that the light yellow region has a cut-off at about 8\,M$_\oplus$ because we exclude massive cores that, given the parameters of the maximum accreted atmospheres, would be subject to runaway accretion. The solid-line contours show how the planet mass and radii change as a function of period for the minimum accreted atmospheres. At the orbital period of 100 days the planets are largely unaffected by the atmospheric loss, and closer to the star the photoevaporation removes atmospheres of the lower-mass planets entirely. For the planets that keep their atmospheres at large periods a decrease in period means little to no change in planet mass. Consequently, for these planets a decrease in period results in an increase in planet radius as atmospheres are hotter and more expanded closer to the star due to stronger stellar irradiation. At small periods the atmospheric loss is significant for all planets, and the trend is reversed.

It is clear from Fig. \ref{fig:MR_relationship} that for planets with radii $R \lesssim 2.3$\,R$_\oplus$ the core accretion of atmospheres in the inner disc predicts larger planet radii than those observed, due to the overestimated envelope mass fractions. The predicted atmospheres are massive enough to populate the region corresponding to planet radii of 1.5\,--\,2\,R$_\oplus$, which is inconsistent with the observed decrease in planet occurrence rates at those radii (sheer grey region). For planets with $R \gtrsim 2.3$\,R$_\oplus$ %this is not immediately obvious, as 
there is a region in which the observed (grey) and the predicted (yellow) mass-radius relationships overlap. %However, 
This overlap corresponds to the (minimum accreted) predicted atmospheres for orbital periods between 20 and 100 days, and a narrow range of short orbital periods (2\,--\,5 days). Notably, even for the minimum accreted atmospheres, the planet radii, at fixed planet mass, are smaller than those observed only for significant high-energy fluxes at orbital periods of less than about 2 days. Taking into account the full range of accreted atmospheres (up to the maximum accreted atmospheres shown in light yellow) further suggests that the predicted atmospheres are typically larger than the atmospheres of the observed sub-Neptunes.

We further compare the predictions of our calculations to sub-Neptune planets with measured masses and radii (taken from \citealt{Wolfgang2016}, excluding the planets where only the upper limit on the mass was known). The observed and the predicted radii and masses are shown in Fig. \ref{fig:MR_ind_planets} in four panels corresponding to four orbital period bins. As in Fig. \ref{fig:MR_relationship}, the yellow and light yellow regions correspond to the predictions from the minimum and maximum atmosphere mass models respectively. To facilitate comparison against the planets that are bare solid cores in each period bin, here we also show the core masses that are predicted to lose their entire atmospheres in a given period bin (yellow lines shown below the dash-dot-dash lines that represent the Earth-like composition mass-radius relationship). Fig. \ref{fig:MR_ind_planets} shows that the masses of the predicted bare cores and the period at which photoevaporation can strip them are largely consistent with those observed. That is, there are no observed planets consistent with the Earth-like composition that are (significantly) more massive than the largest core that the photoevaporation can strip (the upper limit of the yellow line, the predicted bare cores) in each period bin. Fig. \ref{fig:MR_ind_planets} also explicitly demonstrates that for the planets that maintain their atmospheres against the photoevaporation, the predicted planet radii are consistent with or larger than the those observed for the majority of the planets. At long orbital periods (20\,--\,100 days) all planets except one are consistent, within the observational uncertainties, with the predictions (the minimum accreted atmospheres in yellow, the maximum accreted atmospheres in light yellow, and the region in between). At intermediate periods (5\,--\,20 days), about a third of planets that are not bare cores have radii smaller than the predicted radii at the same mass. Finally, at short periods of less than 5 days, there are noticeably 5 planets with radii of $\sim$\,$1.8-2$\,R$_\oplus$ that are neither consistent with the mass-radius relationship of rocky cores, nor with the presence H/He envelopes. This suggests, potentially, that the cores of these planets could contain significant amounts of ice. Still, majority of the short-period planets are consistent with the predictions. Additionally, we reiterate that, while there might be exceptions, the radius distribution of sub-Neptunes is consistent with cores being largely rocky \citep{Owen2017}. Therefore, these results confirm our inference that typically our planets accrete too much gas. In section \ref{sec:discussion} we suggest possible explanations for this result, such as incorrect assumptions of quasi-hydrostatic accretion and negligible heating from planetesimal accretion, or missing mass-loss mechanisms that might act during, or after, disc dispersal.

Overall, the atmospheres accreted in the inner disc are typically in agreement with or larger than those observed, with the exception of planets with significant high-energy fluxes within a very narrow range. This is because core accretion is so efficient that considerable atmospheres can be accreted in the hot and low-density MRI-accreting inner disc and also maintained against photoevaporation.

\section{Discussion and conclusions} \label{sec:discussion}
In this paper we have investigated two aspects of planet formation in the inner disc that is viscously accreting due to the MRI. We considered the earliest phase of planet formation that is evolution of dust particles, and the final phase that is the shaping of the planetary atmospheres. We present simple calculations that include both the accretion and subsequent photoevaporation of close-in super-Earth/mini-Neptune planets. By coupling this processes to perform end-to-end calculations we are able to asses the viability of the \textit{in situ} formation model for close-in planets.

In section \ref{sec:dust}, we consider the evolution of dust grains in the inner disc that is viscously accreting due to the MRI \citep{Mohanty2017arxiv}, that features a local gas pressure maximum at the orbital distance of a few tenths of AU. Taking into account the effect of the MRI-induced turbulence on the dust grain size, we find that fragmentation of particles due to turbulent relative velocities limits the particle size to below few millimeters. As a result, the particles are not efficiently accumulated inside the pressure maximum as hypothesised by \citet{Chatterjee2014,Hu2018}. Regardless of that, as the particles become well coupled to the gas, the radial drift is negated in the inner disc, and the dust-to-gas ratio is enhanced throughout the inner disc. Thus, the local gas pressure maximum might play a lesser role in the \textit{in situ} planet formation than previously thought.

The pressure maximum is, however, still the location of a local density maximum in both gas and dust. We explored if the resulting inner disc structure that is enriched in dust could be susceptible to the onset of the streaming instabilities. This pathway to planetesimals seems to be viable only in a narrow region near the pressure (and density) maximum, for the chosen disc parameters.

The gas is not, however, evolved in this work and effects of the growing amounts of dust onto the MRI have not been taken into account. Dust grains lower the gas ionization levels by absorbing free charges and enhancing recombination rates, as ions recombine on the grains \citep{Draine1987,Ilgner2006}. Charged grains are not themselves well coupled to the magnetic field as they are too massive, and so their presence promotes the non-ideal MHD effects which can suppress the MRI \citep{Sano2000,Ilgner2006,Wardle2007,Salmeron2008,BaiGoodman2009}. The likely result of taking dust effects into account is thus weakened turbulence, and the change of the disc structure in the longer term. The consequences can only be investigated by modelling both the gas and the dust self-consistently. Here, we sketch out a potential scenario by considering the relevant timescales.

Assuming that the steady-state solution of the gas structure (Fig. \ref{fig:gas_structure}) is reached before dust starts affecting the MRI, we expect the dust enhancement of the inner disc to ensue. At a certain dust-to-gas ratio the dust will suppress the MRI, and we expect the levels of turbulence to adapt almost instantly, as the timescale of the magnetic field regeneration is the orbital timescale $t_{\rm orb}$ \citep[e.g.][]{Balbus1991}. With the decreasing levels of turbulence, the dust particle size will rapidly grow due to particle coagulation. The growth due to coagulation happens on the timescales of $\Sigma_{\rm g}/\Sigma_{\rm d} t_{\rm orb}$ \citep[e.g.][]{Brauer2008}, so faster than $10^2 t_{\rm orb}$ if the inner disc is indeed enriched in dust.

The gas disc structure would not change over such short timescales. But, in the absence of the viscous heating due to the MRI, the disc will cool at a timescale of $10^2$\,--\,$10^4 t_{\rm orb}$ \citep[the equilibrium thermal timescale from roughly the pressure minimum to the pressure maximum in the model considered here;][]{Mohanty2017arxiv}. The pressure profile would also follow this timescale, as the vertical hydrostatic equilibrium is established quickly on the orbital timescale ($t_{\rm orb}$). This would likely result in the pressure maximum moving radially inwards.

Concurrently, due to larger particle size and lower turbulent stirring the particles would vertically settle towards the midplane and radially towards the pressure maximum, increasing both the surface density and the midplane bulk density dust-to-gas ratio there. Such formation of a ring of solids could potentially trigger formation of larger bodies, such as planetesimals \citep[as hypothesized by e.g.][]{Chatterjee2014}. The larger particle size and the settling towards the midplane would likely trigger the streaming instability (inwards of the pressure maximum; see Fig. \ref{fig:SI}). However, it is unclear if this could lead to the formation of planetesimals, as a gravitational collapse is unlikely due to the low bulk dust densities and high Roche density in the inner disc. Moreover, as discussed above, the pressure maximum is expected to move radially inwards, and so will the accumulated dust, whereas the Roche density steeply increases inwards. Thus, over time the pressure maximum would need to accumulate significantly more dust to cross the gravitational collapse threshold.

The gas accretion rate and the gas surface density will change slowly in comparison to the above processes, on the long viscous timescale, $\sim$\,$10^3$\,--\,$10^5 t_{\rm orb}$ from the pressure minimum to the pressure maximum in the model considered here \citep{Mohanty2017arxiv}. Suppression of the MRI would lead to increased amounts of gas in the inner disc on this timescale. However, if planetesimals are formed, this would clear the inner disc of the dust grains and the MRI could be induced again, decreasing the gas surface density and moving the pressure maximum outwards. At this stage it is unclear whether these competing processes are balanced in a steady state, or the behaviour of the inner disc is dynamic and quasi-periodic. Such a determination can only be investigated
through self-consistent modelling of dust, gas and the MRI.

At high dust-to-gas ratios the dust also becomes dynamically important, and affects the gas disc structure through the drag backreaction \citep{Nakagawa1986}. The gas rotation profile is then driven towards Keplerian, and as a result the radial gas pressure profile flattens. This, in turn, slows down the radial drift of dust particles. If dust already pile ups in the inner disc due to radial drift being slower than in the outer disc, the dust backreaction amplifies the effect \citep{Drazkowska2016}. In this work, dust enhancement is driven by the dust grains already being completely coupled to the gas in the innermost disc, and thus the effect of dust backreaction would be limited. However, the backreaction would become important if the dust grains grow (e.g. due to the supression of the MRI-induced turbulence discussed above), especially near the pressure maximum. If the dust grains grow in the innermost disc where the pressure gradient is negative, the backreaction would slow down the loss of dust to the star. However, the backreaction would also limit the concentration of dust that can be achieved at the pressure maximum, since it acts to flatten the overall gas pressure profile \citep{Taki2016}.

Furthermore, if super-Earth and mini-Neptune cores indeed form \textit{in situ}, would the inferred low gas surface densities due to the MRI allow them to acquire the observed $0.1-10$\,\% envelope mass fractions? We find that they would. In fact, even after accounting for atmospheric evaporation, the calculated atmospheres tend to overestimate the observed ones.

Could the atmospheric accretion in the MRI-implied disc and the observations be brought into agreement, without invoking an assumption that cores form just before the beginning of disc dispersal \citep[e.g.][]{Ikoma2012,Lee2016}? The calculations shown here do not include several effects which could contribute.

First of all, for core masses smaller than $10$\,M$_\oplus$, the discrepancy could be explained by the ``boil-off'' or core powered mass-loss (\citealp{Owen2016,Ginzburg2018}; see also \citealp{Ikoma2012}), a process in which a planet atmosphere that had not cooled and contracted before the disc dispersal loses its mass. Upon the dispersal the stellar continuum radiation illuminates the planet and launches a Parker wind. The mass loss causes rapid contraction of the atmosphere, and the contraction in turn shuts off the mass loss. Planets that start out with few 10s of percent atmospheres, may be left with 1\,\% after the boil-off. This process precedes the mass loss caused by the stellar high-energy flux considered above, and can operate at larger distances from the star.

%Secondly, we calculate the accreted atmospheres assuming a disc gas surface density corresponding to the location of the pressure maximum in the MRI-implied disc. The gas surface density decreases for orders of magnitude in the inner disc. However, the dependence of the accretion rate on the disc density is weak \citep{Lee2014}, and this would involve a decrease in the accreted envelope mass fraction of only a factor of 2.

%Thirdly, 
Secondly, the scaling relations we use to calculate the accreted atmospheres are derived assuming no sources of heating due to planetesimal accretion, or due to heat deposited in the hypothesized final stage of giant mergers of planetary embryos. The latter could be released for several kyr \citep[e.g.][]{Inamdar2015}, lowering the cooling rate of the atmosphere, and thus allowing less gas to be accreted. Furthermore, the scaling relations assume that the gas inside the planet's Hill sphere is bound and static. Three-dimensional simulations suggest this may not be true and that high-entropy disc material is recycled between the envelope and disc \citep[e.g.][]{Ormel2015,Fung2015,Cimerman2017} potentially modifying the atmospheric cooling rate.  

Finally, if the giant mergers happen between planets, after the disc has fully dispersed, they would likely result in significant atmospheric mass loss. Head-on collisions between Earth/super-Earth-sized planets with few-percent atmospheres can remove tens of percent of the total atmospheric mass \citep{Liu2015,Inamdar2016}. 

Nevertheless, to avoid the runaway accretion for more massive cores, the low gas surface densities the MRI provides are favourable compared to the MMSN environment. Furthermore, gas-poor conditions in this case are provided in a long-lived state, and not in a transient phase \citep[e.g. a transition disc, as proposed by][]{Lee2016}.

In summary, our results support the hypothesis that the MRI-driven accretion in the inner protoplanetary disc could lead to \textit{in situ} planet formation. However, there are several avenues that need to be explored in more detail until we can make quantitative predictions. In particular, the feedback between the enhancement of the dust in the inner disc and the suppression of the MRI, the feedback between the dust enhancement and the gas dynamics, and the role of disc dispersal and boil-off/core-powered mass-loss in shaping the final envelope masses of super-Earths/mini-Neptunes are issues that deserve closer study.

\section*{Acknowledgements}
We thank the reviewer for helpful suggestions that improved the manuscript. We thank Jonathan Tan, Lauren Weiss and Eve Lee for helpful discussions. MRJ is funded by the President's PhD scholarship of the Imperial College London and the "Dositeja" stipend from the Fund for Young Talents of the Serbian Ministry for Youth and Sport. JEO is supported by a Royal Society University Research Fellowship.

%%%%%%%%%%%%%%%%%%%%%%%%%%%%%%%%%%%%%%%%%%%%%%%%%%

%%%%%%%%%%%%%%%%%%%% REFERENCES %%%%%%%%%%%%%%%%%%

% The best way to enter references is to use BibTeX:

\bibliographystyle{mnras}
\bibliography{bibliography}

% Alternatively you could enter them by hand, like this:
% This method is tedious and prone to error if you have lots of references
%\begin{thebibliography}{99}
%\bibitem[\protect\citeauthoryear{Author}{2012}]{Author2012}
%Author A.~N., 2013, Journal of Improbable Astronomy, 1, 1
%\bibitem[\protect\citeauthoryear{Others}{2013}]{Others2013}
%Others S., 2012, Journal of Interesting Stuff, 17, 198
%\end{thebibliography}

%%%%%%%%%%%%%%%%%%%%%%%%%%%%%%%%%%%%%%%%%%%%%%%%%%

%%%%%%%%%%%%%%%%% APPENDICES %%%%%%%%%%%%%%%%%%%%%

%\appendix

%\section{Some extra material}

%If you want to present additional material which would interrupt the flow of the main paper,
%it can be placed in an Appendix which appears after the list of references.

%%%%%%%%%%%%%%%%%%%%%%%%%%%%%%%%%%%%%%%%%%%%%%%%%%

% Don't change these lines
\bsp	% typesetting comment
\label{lastpage}
\end{document}